\documentclass[a4paper,fleqn,usenatbib]{mnras}
\newcommand{\vlos}{$\rm v_{\rm LOS}$}
\newcommand{\feh}{$\rm [Fe/H]$}
\newcommand{\beacon}{{\sc Beacon}}

\usepackage[T1]{fontenc}
\usepackage{ae,aecompl}
\usepackage{graphicx}	

\usepackage{amssymb}

\usepackage{multirow}

\title[Chemo-kinematics of Fornax]{Rotating stellar populations in the Fornax dSph galaxy}
\author[Andr\'es del Pino et al.]{Andr\'es del Pino$^{1}$\thanks{E-mail: \mbox{adpm@camk.edu.pl}},
Antonio Aparicio$^{2,3}$, Sebastian L. Hidalgo$^{2,3}$ and 
 Ewa L. {\L}okas$^{1}$\\
$^1$Nicolaus Copernicus Astronomical Center, Polish Academy of Sciences, Bartycka 18, 00-716 Warsaw, Poland\\
$^2$Instituto de Astrof\'\i sica de Canarias, Calle V\'\i a L\'actea s/n,
E-38200 La Laguna, Tenerife, Spain\\
$^3$Departamento de Astrof\'\i sica, Universidad de La Laguna, Avda. Astrof\'isico Fco. S\'anchez s/n, E-38206 La
Laguna, Tenerife, Spain}

\begin{document}


\pagerange{\pageref{firstpage}--\pageref{lastpage}} \pubyear{2016}

\maketitle

\label{firstpage}

\begin{abstract}
We present a novel analysis of the internal kinematics of the Fornax dwarf spheroidal galaxy. Our results are
based on the largest sample of spectroscopic data for Fornax stars presently available ($> 2500$ stars), for which we
have chemical and kinematic information. We introduce new software, \beacon, designed to detect chemo-kinematic
patterns among stars of different stellar populations using their metallicity and velocity along the line of sight.
Applying \beacon\ to Fornax we have detected non-negligible rotation signals around main optical axes of the galaxy,
characteristic for a triaxial system partially supported by rotation. The dominant rotation pattern is relatively strong
($\sim 12$ km s$^{-1}$), but the galaxy also shows additional weaker albeit complex rotation patterns. Using the
information available from the star formation history of Fornax we have also derived the average age of the
different chemo-kinematic components found by \beacon, which has allowed us to obtain its
kinematic history. Our results point to a possible major merger suffered by Fornax at redshift $z\sim1$, in agreement
with previous works.

\end{abstract}

\begin{keywords}
galaxies: evolution -- galaxies: Local Group -- galaxies: dwarf -- galaxies: kinematics and dynamics
-- galaxies: interactions  -- galaxies: individual: Fornax
\end{keywords}

\section{Introduction}
\label{introduccion}

Dwarf galaxies comprise the most common type of galaxies in the Universe, however even in the Local Group the details
of the internal kinematics of many of them remain unknown. This is specially true for the most recently discovered
ultra faint dwarfs, but also for many of the well known classical dwarf spheroidal (dSph) galaxies. There are several
reasons for this; some of them more technical or observational, others more conceptual.

The first of the observational issues is related to their relatively large distances, up to $(m-M)_0 \sim 24$ for many
of the M31 satellites \citep{McConnachie2012}. Observations of these objects thus require the use of large
telescopes and even then obtaining e.g. the proper motions of their stars remains extremely difficult. The second main
technical problem is the fact that dSph galaxies retain very little gas. Consequently, the spectra of a large number of
their stars must be obtained in order to reliably study their kinematics. This has posed a technological challenge
and only in the most nearby galaxies a sufficient number of stars have been measured to provide solid results.

Some of the classical dSphs have been repeatedly studied during the last decade. In most of them, no obvious signs of
rotation were found \citep{Wilkingson2004, Munoz2005, Koch2007a, Koch2007b, Walker2009a, Wheeler2015}. In the case of the Milky Way
(MW) satellites, the evidence for gradients in the line-of-sight velocity (\vlos) along the optical axes is most
convincing for Fornax, Sculptor and Sextans dSphs \citep{Battaglia2008a, Battaglia2011, Amorisco2012}.
However, the absence of reliably measured proper motions for some of these galaxies makes it unclear whether these
gradients are the intrinsic rotation signal or rather produced by geometrical effects \citep{Walker2008, Strigari2010}. Therefore,
dSph galaxies have been generally considered to be pressure-supported systems with little or no rotation.

From the cosmological point of view, dSphs are likely to be examples of the evolved state of late-type dwarfs,
transformed into spheroids as a result of tidal stirring, ram pressure and other environmental effects \citep[][and
references therein]{Mayer2010}. If this scenario is correct, some residual rotation signal should be present in
these galaxies. Could these rotation signals have escaped detection due to the large velocity dispersions of
these systems? In the case they do not rotate at all, how did they acquire their spheroidal shapes and are not simply
spherical as globular clusters?
These questions encouraged us to develop \beacon, a code designed specifically for finding chemo-kinematic patterns in
resolved stellar systems. In this paper, we test the capabilities of \beacon\ and apply it to the Fornax dSph galaxy.

With about $\sim$2500 likely member stars observed, Fornax represents the state-of-the-art system in terms of
spectroscopic observations in dSph galaxies. Fornax is, after the Sagittarius dSph, the largest and most luminous of
the dSph MW companions, with a core radius of $16.6^\prime \sim$ 550 pc. It is located at a distance of 136$\pm$5 kpc
\citep*{Mackey2003, Greco2007, Greco2009, Tammann2008, Poretti2008}, and its principal baryonic component is stellar,
with an ambiguous detection of $\rm HI$ \citep*{Bouchard2006} possibly associated with the MW. The dynamical mass
within the observed half-light radius has been estimated to be $\rm 5.6 \times 10^7 M_\odot$ \citep[][and references
therein]{McConnachie2012}, while the total mass (assuming mass follows light) is $\rm 1.6 \times 10^8 M_\odot$
\citep{Lokas2009}. The observational properties of the Fornax dwarf are summarized in Table~\ref{tab:Fornax_Glance}.

Fornax is a remarkable object to study. Together with Sagittarius, it is the only dSph among MW companions harbouring a system of  
globular clusters. Its
star formation history (SFH) is long-lasting and complex, showing strong spatial gradients between the centre and the
outskirts of the galaxy \citep{deBoer2012, delPino2013}. This can be also viewed as a clear spatial segregation
between its different stellar populations \citep{delPino2015}. Moreover, its young stellar populations are not
symmetrically distributed in the galaxy, showing stellar clumps and structures not aligned with the optical axes.
The most famous ones are the shell-like overdensities located at $17^\prime$ and 1.3$^{\circ}$ from the
centre, although the latter has been recently claimed to be in fact an overdensity of background galaxies
\citep{Bate2015}. Its internal dynamics also appears to be rather complex, leading to diverse results and
interpretations concerning the past of the galaxy \citep{Walker2009a, Amorisco2012, Hendricks2014}.

All these observational facts have stoked a debate about the possibility that Fornax suffered a merger. Some authors
claim that these stellar substructures are the result of a past merger \citep{Coleman2004, Coleman2005, Amorisco2012,
Yozin2012, delPino2015}. On the other hand, \citet{deBoer2013} claim that these clumps are more likely the result
of a quiet infall of gas previously expelled from Fornax during its star formation episodes.

The large amount of data available for Fornax together with its apparently complex past have made of this galaxy a
perfect target for testing \beacon. In this paper, we applied the code on the largest ensemble of
spectroscopic data for Fornax stars available to date. In Section~\ref{Cap:Data} the data set is presented and we
explain the internal calibration between catalogues we performed. Section~\ref{Cap:Method} describes \beacon. In
Section~\ref{Cap:Results} we present the results of its application to Fornax. In Section~\ref{Cap:Consistency},
consistency tests and their results are described. Our results are discussed in Section~\ref{Cap:Discussion}. Finally, a
summary and the main conclusions of the paper are presented in Section~\ref{Cap:Conclusions}.

\begin{table}
\begin{minipage}{\linewidth}
\caption{Observational parameters of the Fornax dSph.}
\label{tab:Fornax_Glance}
\begin{tabular}{@{}lcc}
\hline
\hline
Quantity & Value & Reference\footnote{(1) \citet{Irwin1995}; (2) \citet{Mateo1998}; (3) \citet{van_den_Bergh1999}; (4)
\citet{McConnachie2012}.}\\
\hline
RA, $\alpha$ (J2000.0) &  2h 39$^\prime$ 53.1$^{\prime\prime}$ & (3)\\
Dec, $\delta$ (J2000.0) & -34$^{\circ}$ 30$^\prime$ 16.0$^{\prime\prime}$ & (3)\\
Heliocentric velocity (km s$^{-1}$) \hspace{20pt}& 55.3$\pm$0.1 & (4) \\
Ellipticity, $e$ & $0.30\pm 0.01$ & (1)\\
Position angle ($^\circ$) & 41$\pm$6 & (1) \\
Core radius ($^\prime$) & 13.8$\pm$0.8 & (2)\\
Tidal radius ($^\prime$) & 71$\pm$4 & (2)\\
\hline
\end{tabular}
\end{minipage}
\end{table}

\section{Data}\label{Cap:Data}

There is a large number of spectroscopic measurements for red giant branch (RGB) stars in Fornax. These comprise data
collected with a variety of instruments, resolution and scientific aims. Therefore, not all available catalogues
provide equally comprehensive information, nor of the same quality. In principle, to get the most complete information
about each star (e.g. $\alpha$-element abundances) it would be desirable to have high resolution
spectroscopy with a good signal-to-noise (S/N) ratio. Since this is not the case for a vast majority of the observed
stars in Fornax, it is then preferable to have more basic information but for as many stars as possible. For this
reason, we have based our study only on the metallicity (\feh) and line-of-sight velocities (\vlos), which are commonly
provided by most of the available spectroscopic catalogues for Fornax. We decided to collect this information from the
catalogues with the largest number of observed stars in Fornax \citep{Pont2004, Battaglia2006, Battaglia2008b,
Walker2009a, Kirby2010, Letarte2010}.

\subsection{The calcium triplet}

The infrared Ca II triplet lines (CaT) have been repeatedly measured for more than a 1000 stars in Fornax
\citep{Pont2004, Battaglia2008b, Kirby2010, Letarte2010}. In terms of metallicity these works provide rather consistent
results \citep[see][for further information]{delPino2013}. Notwithstanding, the different metallicity scales used in
each study make a direct comparison between catalogues difficult. In order to make all metallicities compatible with
each other we have rescaled each catalogue to the metallicity scale of \citet{Battaglia2008b} (B08 hereafter) using the
\citet{Starkenburg2010} calibration. This scale was preferred since the age-metallicity relation (AMR) obtained using
their data is in good agreement with the one obtained from deep VLT photometry \citep{delPino2013}.

\citet{Letarte2010}, provide \vlos ~as well as the associated errors for their observed stars in their table A.1. In the
case of B08 and \citet{Kirby2010}, these quantities were kindly provided by the authors. In order to obtain \vlos ~for
\citet{Pont2004} stars, we cross-matched these with the \citet{Walker2009a} (W09 hereafter) catalogue. Only good matches
were saved in the final catalogue, keeping the recalibrated metallicities from \citep{Pont2004} and the \vlos ~from
W09. In total, we used 1369 stars from these catalogues.

\subsection{The magnesium triplet}
\label{introduccion:magnesium}

The W09 spectra sample the $\sim 5160$ \AA{} region, which contains the magnesium triplet absorption lines. The
metallicities obtained in this work differ significantly from the rest of the catalogues. We believe this is an effect caused by an incomplete sky subtraction in some of their measurements, since their linear model works correctly in their
calibration globular clusters. In order to be able to compare the W09 stars with the rest of the sample, we recalculated
their metallicities using the \citet{Starkenburg2010} empirical model with their $\rm \Sigma Mg$ composite index ($\rm
\langle \Sigma Mg \rangle$ when several measurements were available). The \citet{Starkenburg2010} model defines the
metallicity of a star, $\rm [Fe/H]_{cal}$, as:
\begin{eqnarray}
\rm [Fe/H]_{cal} &=& a + b\ (V-V_{\rm HB}) + c\ \Sigma {\rm Mg}   \label{Eq:Calibration} \\
                 &+& d\ \Sigma {\rm Mg}^{-1.5} + e\ \Sigma {\rm Mg} (V-V_{\rm HB})  \nonumber
\end{eqnarray}
where $V$ is the magnitude of the star, $V_{\rm HB}$ is the magnitude of the horizontal branch (HB), and $\Sigma {\rm
Mg}$ is the composite index of the magnesium triplet lines of the star.

Model parameters $(a, b, c, d, e)$ were obtained through least squares minimization method.
The presence of numerous local minima in the $\chi^2$ function makes the process strongly
dependent on the starting guess
parameters. To overcome this, we minimized a penalized $\chi^2$ function of the form:
\begin{equation}
\chi^2 = \frac{(1-ks)^2 (\rm [Fe/H]_{cal} - [Fe/H]_{CaT})^2}{\rm [Fe/H]_{cal}^2 \
[\sigma^2(\rm [Fe/H]_{cal}) + \sigma^2(\rm [Fe/H]_{CaT})]} \label{Eq:Chi}
\end{equation}
where $ks$ is the Kolmogorov-Smirnov statistic resulting from comparing the $\rm [Fe/H]_{cal}$ and $\rm
[Fe/H]_{CaT}$ metallicity distributions. $\rm [Fe/H]_{cal}$ is the metallicity derived using
equation~(\ref{Eq:Calibration}) and $\rm [Fe/H]_{CaT}$ is the metallicity of the star obtained through CaT
spectroscopy. Their uncertainties are $\sigma(\rm[Fe/H]_{cal})$ and $\sigma(\rm[Fe/H]_{CaT})$, respectively. We
minimized expression~(\ref{Eq:Chi}) using the stars present both in the W09 and B08 catalogues, adopting metallicities
from the latter as $\rm [Fe/H]_{CaT}$. In order to avoid local minimums we proceeded in two steps. First, we created
an $N \times n$ array covering a realistic domain of the $N$ sampled parameters with $n$ values for each parameter.
Once the whole domain of starting points has been scrutinized and the total minimum has been found, its $N$ coordinates
are used as a starting point for a more refined least squares minimization of (\ref{Eq:Chi}).

To avoid stars with large errors, or with low S/N, only those with $\sigma({\rm \Sigma Mg}) < 0.08$ were considered.
The errors in the metallicities were obtained by propagating the errors in all the quantities implied in the
calculation: $V$, $V_{\rm HB}$ and $\sigma({\rm \Sigma Mg})$. W09 only provide the $\sigma(\rm \Sigma Mg)$
uncertainty. The other uncertainties were obtained by cross-matching the W09 stars with a photometric catalogue
covering the whole body of the galaxy \citep[see][for further information]{delPino2015} and assigning to each star its
$V$ magnitude and its corresponding error $\sigma(V)$ from the photometry catalogue. This was done after ensuring
that the $V$ magnitudes from both catalogues were compatible within errors. The uncertainty in $V_{\rm HB}$ was set to
$\sigma(V_{\rm HB}) = 0.05$. This value is the standard deviation of 50 measurements of $V_{\rm HB}$ over the
photometry catalogue in different non-overlapping regions covering small areas within the whole region spanned by W09
stars.

\begin{figure}
\begin{center}
\includegraphics[scale=0.6]{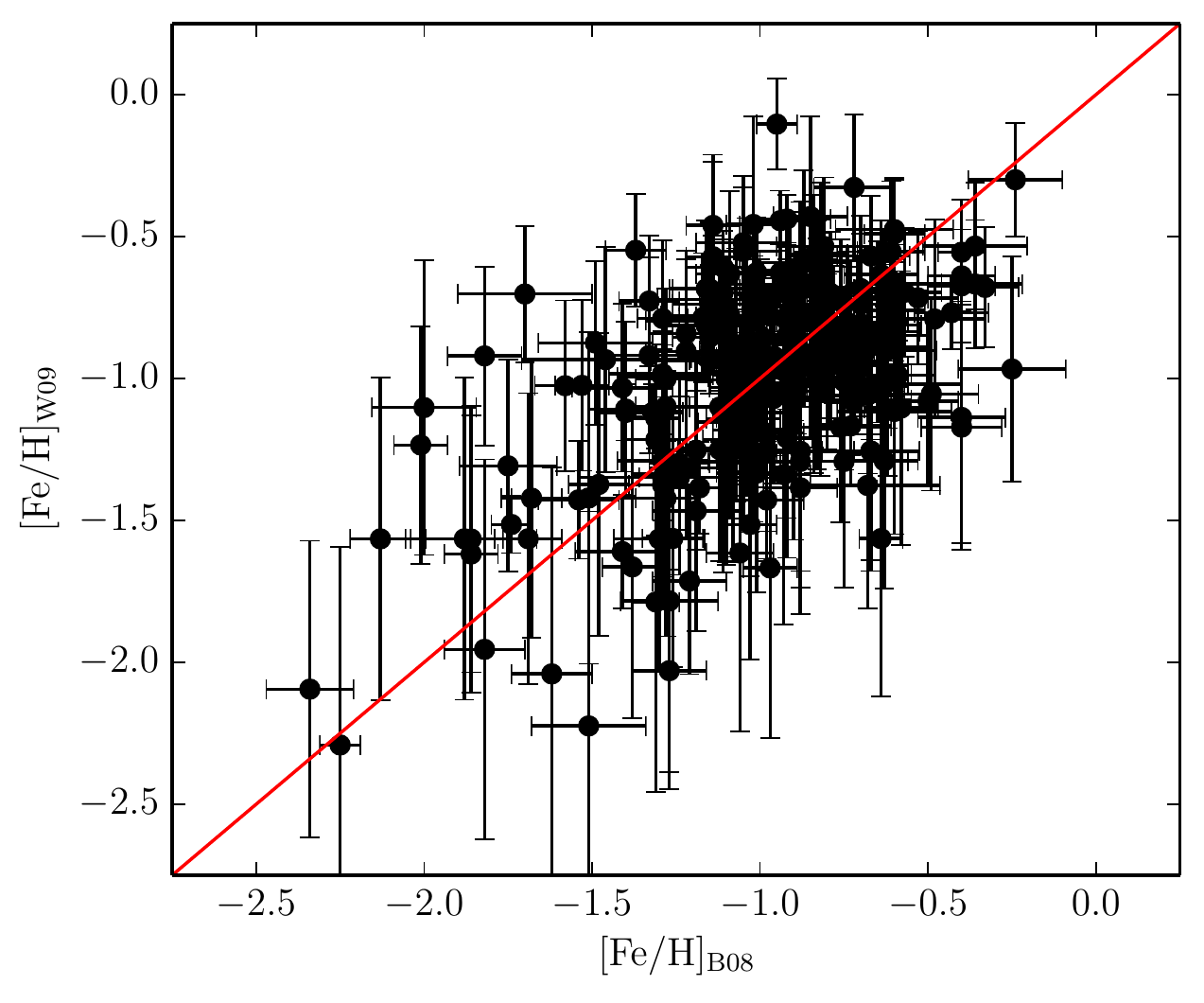}
\caption[Comparison between metallicity of common stars]{Comparison of metallicities of stars common in the W09 and B08
catalogues after the internal calibration. W09 metallicities have been recalibrated to agree with B08 ones using the
\citet{Starkenburg2010} model. The red line indicates the one-to-one correspondence between the two catalogues.}
\label{fig:Z_calibration}
\end{center}
\end{figure}

\begin{figure}
\begin{center}
\includegraphics[scale=0.6]{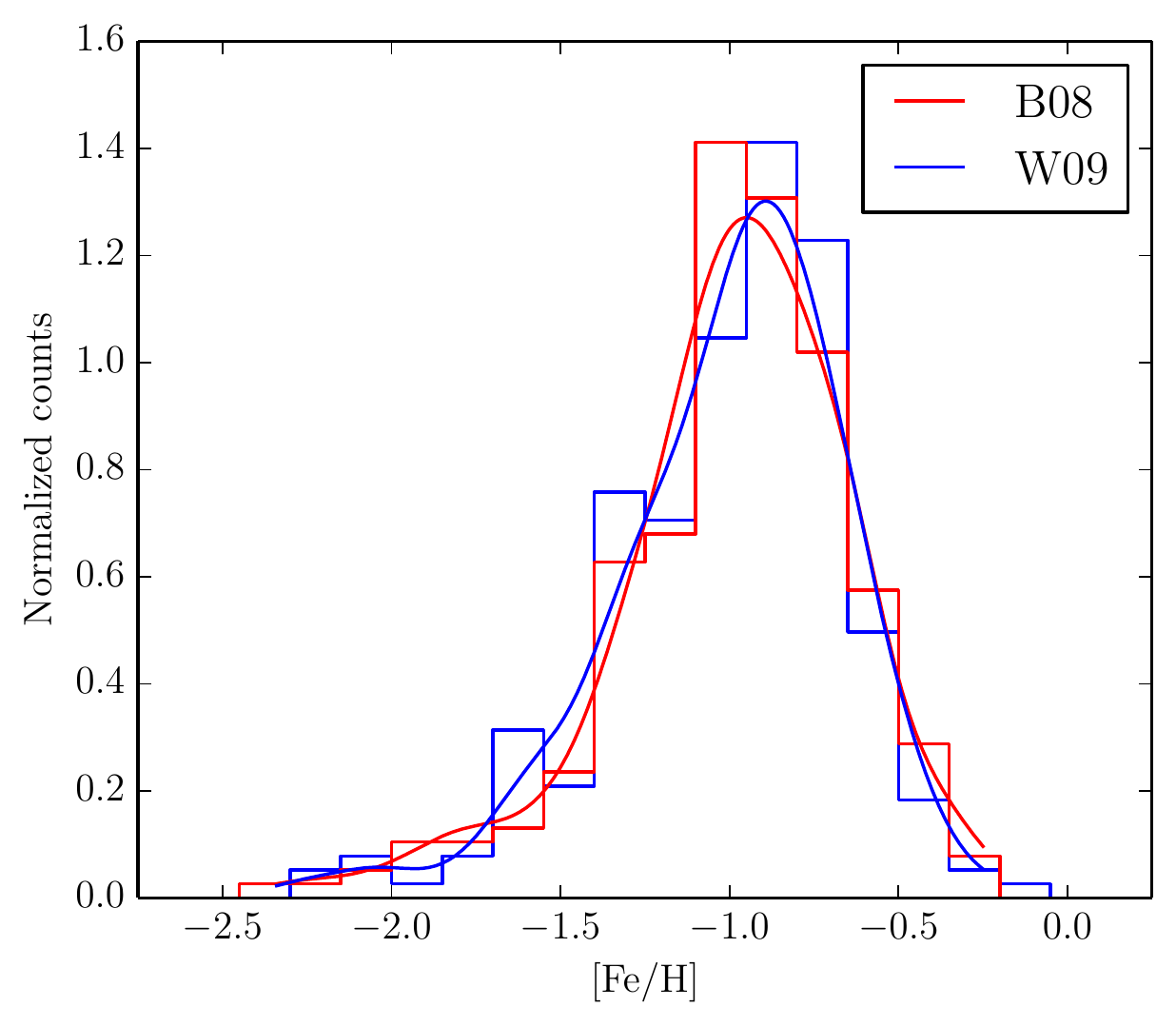}
\caption[Comparison between metallicity distributions after the internal calibration]{Comparison between metallicity
distributions of W09 and B08 stars after the internal calibration. Metallicity distribution of W09 is shown in blue and
the one of B08 in red. W09 metallicities have been recalibrated to agree with B08 ones using the \citet{Starkenburg2010}
model. Curves represent the fitted distribution functions of metallicities from W09 (blue) and B08 (red).}
\label{fig:Z_distro}
\end{center}
\end{figure}

After the calibration, the metallicity distribution of W09 follows closely the one of B08 so the disagreement seen
when considering particular stars becomes smaller when we consider a statistically significant number of stars. In
Figure~\ref{fig:Z_calibration} we show a comparison between the metallicities of the stars common to both catalogues
after the internal calibration. Figure~\ref{fig:Z_distro} shows the metallicity distribution of the same stars from W09 and B08 (also after
the internal calibration).

\begin{figure}
\begin{center}
\includegraphics[ scale=0.6]{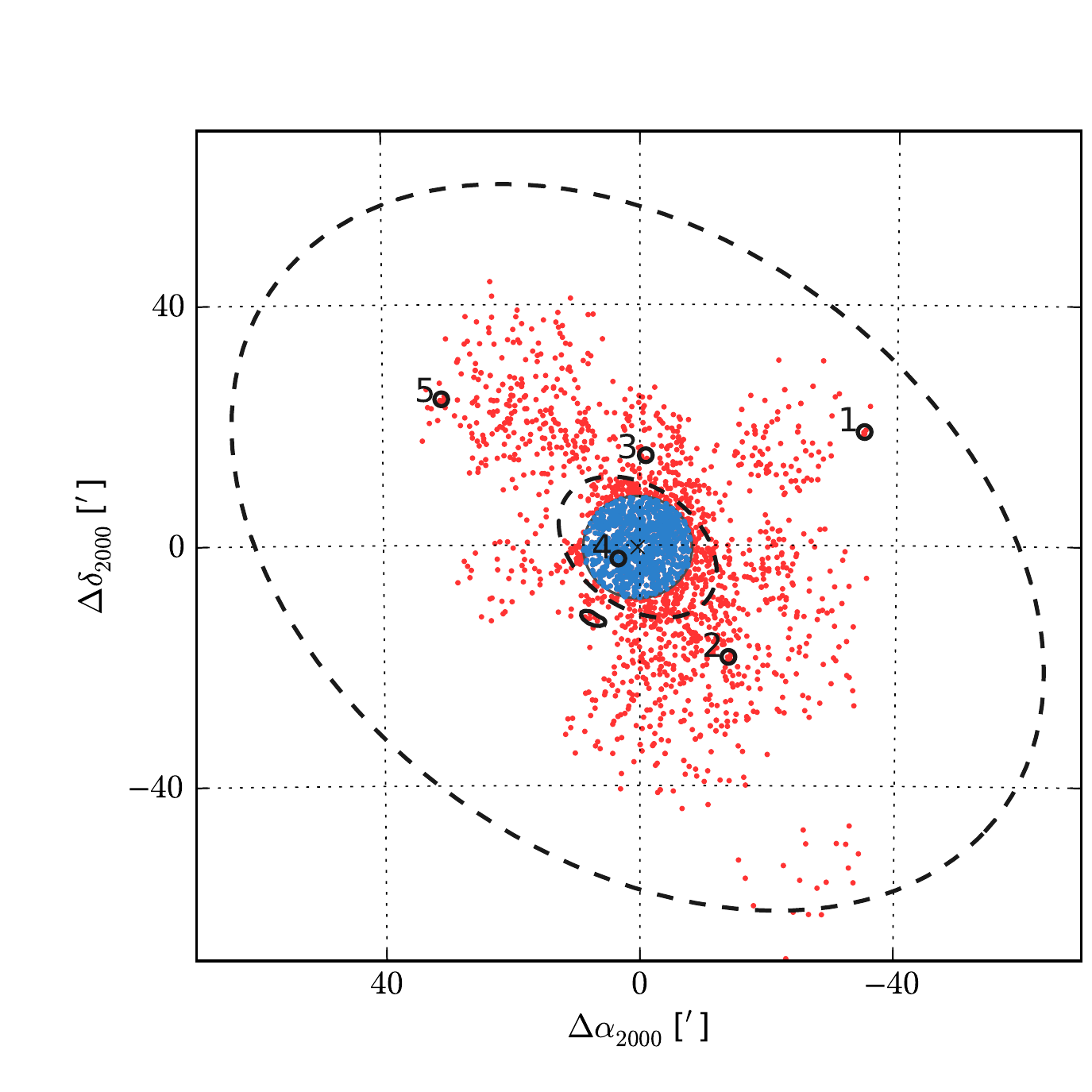}
\caption[Stars used for the chemo-kinematic analysis]{Stars used for the chemo-kinematic analysis of Fornax. These were collected from \citet{Pont2004, Battaglia2006, Battaglia2008b, Walker2009a, Kirby2010} and  \citet{Letarte2010}. The positions of the stars are represented by dots. Blue dots indicate a sub-sample of stars, selected for a consistency test (see Section~\ref{Cap:Consistency}). These are all the stars of the sample lying within a circle of radius $r = 7.7^\prime \sim 300$ pc located at the centre of the galaxy. The dashed ellipses correspond to the core and the tidal radius. The black open circles
show the positions of five known globular clusters in Fornax and are labelled accordingly. The position of the inner
shell found by \citet{Coleman2004} is marked with a small ellipse at $\Delta \alpha_{2000} \sim 8^{\prime}$ and $\Delta
\delta_{2000} \sim -13^{\prime}$.}
\label{fig:datasets}
\end{center}
\end{figure}

\begin{figure*}
\begin{center}
\includegraphics[ scale=0.7]{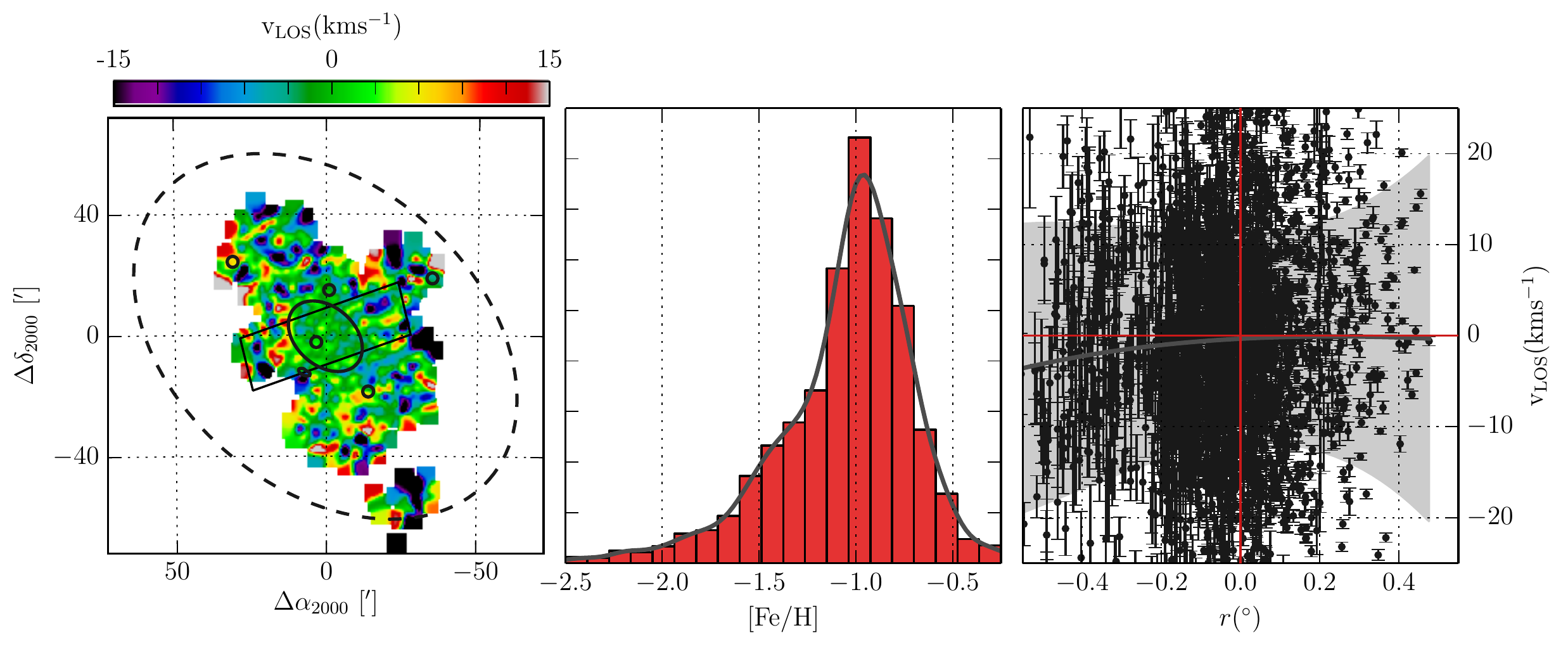}
\caption[Stars used for the chemo-kinematic analysis]{The map of line-of-sight velocities, \vlos\, for the 2562 stars
used in the present work. The left panel shows the \vlos\ as a function of position. The core and the tidal radius are
represented by dashed ellipses. Empty circles indicate the positions of globular clusters in Fornax. The middle panel
plots the metallicity distribution of the stars shown in the left panel. The right panel shows the velocities of the
stars along the stripe marked in the left panel. The average velocity along the stripe is shown by a grey line, while
its dispersion is represented by the shaded area. Voronoi tessellation techniques were used in order to obtain a constant
number of stars per grid cell.}
\label{fig:rawmaps}
\end{center}
\end{figure*}

A total of 1193 stars from the W09 sample were included in the final catalogue, which contains the information about
the position, metallicity, and \vlos ~for 2562 stars. The relative positions of the stars in the galaxy are shown in
Figure~\ref{fig:datasets}. In Figure~\ref{fig:rawmaps}, we show the \vlos ~map of the whole sample, together with its
metallicity distribution and a stripe showing the maximum detected velocity gradient ($\sim 4$ km s$^{-1}$ deg$^{-1}$) 
along the body of Fornax. Note that the direction of this velocity gradient, PA = 112 $\pm$ 9 deg, is 
compatible with the direction of the proper motion vector of Fornax derived by \citet{Piatek2007}, PA = 127$\pm$4.1 deg 
or \citet{Walker2008}, PA = 120 deg. We believe that this velocity gradient is in fact the projection along the 
line-of-sight of the transversal proper motion of Fornax and not an intrinsic velocity gradient.
In order to have more statistically reliable results in the velocity maps,
we have made use of Voronoi tessellation techniques. With this method, an adaptive grid is used instead of a regular one, adjusting the sizes of the cells to keep a constant number of stars within the cells. Areas
where there were less observed stars, are covered by larger
grid cells. This method was simultaneously used to clean the
sample from outliers. The tessellation was performed in an
iterative way until convergence, removing from every cell stars
lying outside the 3$\sigma$ of the velocity distribution within the
cell.

\section{BEACON: a method to find chemo-kinematic patterns}\label{Cap:Method}

Disentangling chemo-kinematic patterns among the stars of a dSph is difficult. There have been numerous attempts of
detecting rotation, generally concluding that no strong rotation signal can be found in these systems.
Nevertheless, significant rotation signatures have been reported for some dSph
galaxies in the literature. Using \vlos ~measurements for $470$ RGB stars \citet{Battaglia2008a} found a small
velocity gradient of $7.6^{+3.0}_{-2.2}$ km s$^{-1}$ along the projected major axis of the Sculptor dSph galaxy.
Furthermore, \citet{Lewis2007} and \citet{Fraternali2009} claimed to have detected a regular rotation around the minor
axis in the Cetus and Tucana dSph galaxies with a velocity difference of $\sim 8$ km s$^{-1}$ and $\sim 16$ km
s$^{-1}$ respectively. This conclusions were however based on the \vlos\ measurements for only a few tens of RGB stars
in each galaxy which is not enough to derive solid results. More recently, \citet{Amorisco2012} found a small rotation
signal in Fornax using the W09 data set. The small number of dSphs where dynamical patterns have been found
together with their relatively weak rotation signals have supported the general idea of dSphs being pressure
supported systems. Therefore, their elongated shapes were usually interpreted as a result of tidal interactions with
their larger host galaxies.

However, our technological limitations could be, at least in part, responsible for these results. We are only able
to obtain spectra of a limited number of individual RGB stars, therefore sampling a very small fraction of the total
stellar population of these galaxies. Due to this poor sampling a weak rotation signal could well be hidden in a
relatively large velocity dispersion. These observational limitations make it necessary to implement new software
techniques to help us find dynamical patterns in these galaxies.

\beacon~is a new software tool developed to find groups of stars with similar chemo-kinematic properties. Its
scientific objective is to detect possible rotation patterns in resolved stellar populations. It has been written in
the Python language and we are currently working to make the code publicly available to the scientific community.

\subsection{Basic Definitions}\label{Subcap:deffinitions}

From the point of view of dynamics, galaxies are $N$-body systems where each star can be defined by its position and
velocity. Therefore, the internal dynamics of a galaxy with $N$ stars can be described by a state vector, $\mathbf{x}
\in \mathbb{R}^{6N}$, where $\mathbf{x} = (\mathbf{r}_1 \ldots \mathbf{r}_N, \mathbf{v}_1 \ldots \mathbf{v}_N)$ and
each $\mathbf{r}_1 \ldots \mathbf{r}_N, \mathbf{v}_1 \ldots \mathbf{v}_N \in \mathbb{R}^{3}$.

The most important issue that observational astronomers have to deal with is the fact that observations only provide
information about the projection of this state vector in 3 out of 6 coordinates and obviously not for all the $N$
stars of the system but only for a small observed fraction, $n$. These 3 coordinates usually are the classical
coordinates RA, Dec, and \vlos, which are, respectively, the projection of the position on to the celestial sphere and
the projection of the velocity on to the line of sight: $\mathbf{x}^{p} = (\mathbf{r}_1^{p} \ldots \mathbf{r}_n^{p},
\mathbf{v}_1^{p} \ldots \mathbf{v}_n^{p})$ where the superscript $p$ stands for `projected'.

These pieces of information are insufficient to unequivocally determine the dynamical state of a galaxy, but allow us to
make some inquiries through assumptions or models. The first and most important model to be considered is the one
assuming that the system should be in virial equilibrium and is kinematically supported. Stars should be orbiting
around the centre of mass (CM) of the galaxy with trajectories that follow a common rotation pattern (rotation
supported systems) or are randomly oriented (pressure supported).

Disentangling kinematic patterns becomes difficult when using only the projection of the state vector of a reduced
sample of stars. To improve our knowledge about the state of each star, one more assumption can be imposed: we can
suppose that stars showing similar kinematic patterns should also have similar chemical composition. The latter
assumption is based on the idea that stars would conserve the kinetic properties of the gas cloud from which they were
formed. In the following, we will concentrate on distinguishing possible rotation patterns using a state vector
defined as $\mathbf{\Phi} = (\mathbf{r}_1^{p} \ldots \mathbf{r}_n^{p}, \mathbf{v}_1^{p} \ldots \mathbf{v}_n^{p}, {\rm
Z_1} \ldots {\rm Z}_n)$, were ${\rm Z}_i$ are the metallicities of the stars.

\subsection{\beacon~in a nutshell}\label{Subcap:BEACON}

\subsubsection{Outline of the procedure}

Finding streams of material with similar chemo-kinematic patterns can be mathematically expressed as a hierarchical
clustering problem. This can be addressed via different approaches, for example the Bayesian techniques
of \citet{Amorisco2012}. In the present work, we propose a completely different strategy. We have developed a software
suite called \beacon~which has been optimized for joining together groups of stars with similar positions, velocities
and metallicities. The core of \beacon~is based on the {\sc Optics} algorithm by \citet{Optics1999} and can be divided
into the following steps:
\begin{itemize}
 \item[1.] Building the state vector $\mathbf{\Phi}$;
 \item[2.] Clustering process of $\mathbf{\Phi}$;
 \item[3.] Analysis of clusters;
 \item[4.] Recovering the original coordinates of clustered stars;
 \item[5.] Uniqueness of solutions and groups creation.
\end{itemize}

In the following we provide a description of each step:

1. Building the state vector $\mathbf{\Phi}$: Having measurements for $n$ stars, we have to create a convenient state
vector which represents our sample. In this vector, we may include any coordinate providing information of interest
about the state of the stars: positions, \vlos\ and metallicities. Positions and velocities of stars are transformed
into the galaxy CM rest frame of reference (RA$^{\rm CM}$, Dec$^{\rm CM}$, $\rm v^{\rm CM}_{\rm LOS}$) and for convenience
position coordinates are converted into polar coordinates defined as $(r, \theta)$. Since the orbits followed by the
streams of stars do not have to be perfectly circular, $r$ should be a free parameter and should not to be included in
$\mathbf{\Phi}$. The latter decision changes the possible usages of \beacon. For example, including $r$ would make
\beacon~more sensitive to streams or clumps of stars more concentrated in the space defined by $\mathbf{\Phi}$, but less
effective in detecting more general chemo-kinematical patterns. Since our objective is to study general rotation
patterns in the galaxy, we did not include $r$ in $\mathbf{\Phi}$.

All quantities can be standardized before the clustering process. Standardization of input variables will normalize the
relative weight of each of them during clustering. This allows us to avoid undesirable effects related to the
dynamical range covered by the variables involved in the clustering process. For example, we wish to avoid the
situation when the \vlos, which can take values from the range of almost $\pm 25$ km s$^{-1}$ weights much more than the
metallicity, \feh, which only covers $\sim 0.008$ in dex scale. In case it was required, standardization to standard
deviation, dynamical range and the variance of variables are all included in the code.

After the process, the state vector would have the explicit form of:
\begin{equation}
\mathbf{\Phi} = \left[ \begin{array}{ccc}
                         \theta_1 &  {\rm v}_{{\rm LOS},1} & Z_1 \\
                         \theta_2 &  {\rm v}_{{\rm LOS},2} & Z_2 \\
                         \vdots   &   \vdots   & \vdots \\
                         \theta_n &  {\rm v}_{{\rm LOS},n} & Z_n \\
                       \end{array} \right]
\end{equation}
where each $\theta_i$ is the angular coordinate of the $i^{\rm th}$ star with respect to the CM and the optical minor
axis of the galaxy and each ${\rm v}_{{\rm LOS},i}$ is measured with respect to the ${\rm v}^{\rm CM}_{\rm LOS}$.

2. Clustering process of $\mathbf{\Phi}$: \beacon~will then create a new state vector, symmetrical in relation to
$\mathbf{\Phi}$ in coordinates and velocities with respect to the centre of the galaxy. Both vectors are stacked into a
unique vector $\mathbf{\Theta}$ containing $2n$ stars:
\begin{equation}\mathbf{\Theta} = \left[ \begin{array}{ccc}
                         \theta_1 &  {\rm v}_{{\rm LOS},1} & Z_1 \\
                         \vdots   &   \vdots   & \vdots \\
                         \theta_n &  {\rm v}_{{\rm LOS},n} & Z_n \\
                         \theta_1+\pi \pmod{2\pi} &  -{\rm v}_{{\rm LOS},1} & Z_1 \\
                         \vdots   &   \vdots   & \vdots \\
                         \theta_n+\pi \pmod{2\pi} &  -{\rm v}_{{\rm LOS},n} & Z_n \\
                         \end{array} \right]
\end{equation}
This $\mathbf{\Theta}$ vector is the input for the {\sc Optics} algorithm. The use of $\mathbf{\Theta}$ instead of
$\mathbf{\Phi}$ allows us to cluster stars with similar symmetrical positions ($\theta$ and
$\theta +\pi$), which are moving with similar velocities but in the opposite direction (\vlos ~and $-$\vlos).

3. Analysis of clusters: Once the stars have been clustered, \beacon~searches in each cluster for stars located at both
sides of the CM of the galaxy. We define $\epsilon$ as the ratio of the numbers of stars at one and at the other
side of the CM. If $\epsilon$ is greater than a certain critical quantity, $\epsilon_{\rm c}$, the cluster is
catalogued as a possible circular or both side stream (BSS). On the other hand, if all stars are on one side of
the CM ($\epsilon_{\rm c}=0$), the cluster is catalogued as a possible one side or non-circular stream (OSS).
Clusters with $0 < \epsilon < \epsilon_{\rm c}$ are dismissed.

4. Recovering the original coordinates of clustered stars: The original coordinates, containing the real physical data,
of the stars clustered in $\mathbf{\Theta}$ are recovered from $\mathbf{\Phi}$.

5. Uniqueness of solutions and groups creation: In this last step, \beacon~checks whether common stars between clusters
exist, and applies uniqueness criteria. Two alternative uniqueness criteria are implemented: \textit{all elements} or
\textit{any element}. The former considers two clusters as different groups if they differ by at least one star. The
latter criterion considers clusters sharing any stars as coming from the same group and merges them all into one
group.

\subsubsection{Additional parameters}

Apart from the observational data, some additional information is required in order to perform the clustering. Such
information is used to create the reference frame centred on the CM of the galaxy (\textit{galaxy parameters})
and to control the clustering process (\textit{clustering parameters}).

In order to work properly, \beacon~needs the aforementioned parameters as input. The \textit{galaxy parameters} are the
coordinates of the CM. More data can be provided, but only the position and the \vlos ~of the CM (RA$^{\rm CM}$,
Dec$^{\rm CM}$, ${\rm v}^{\rm CM}_{\rm LOS}$) are strictly needed. These coordinates allow \beacon~to create the reference
frame centred on the CM, which is necessary in order to detect rotation patterns in the galaxy. The \textit{clustering
parameters}, on the other hand, are a collection of parameters defining the clustering criteria. Some of them are the
\textit{standardization method}, the \textit{uniqueness criteria} and the \textit{minimum cluster size} (MCS). These
have to be tuned for the specific galaxy under study and depend for example on the number of sampled stars, the
completeness of the sample and its spatial coverage.

\subsection{The case of Fornax: optimal parameters}\label{Cap:Method_in Fornax}

For the analysis of Fornax, we initially adopted the galaxy parameters listed in Table~\ref{tab:Fornax_Glance}. 
On the other hand, the \textit{clustering parameters} were fine-tuned through several tests. In these tests we analyzed the solutions of \beacon~as a function of variations in the clustering parameters. 

The \textit{standardization method} affects the relative importance of each coordinate during the clustering. In
principle, nothing suggests that any coordinate should have a larger weight than any other in the clustering process.
Therefore we applied a standard deviation standardization. This method assigns the most balanced weight to the three
coordinates ($\theta$,  \vlos, and $Z$). Also the \textit{uniqueness criteria} were set to `any element', in order to
avoid duplicities in stars.

\begin{figure}
\begin{center}
\includegraphics[ scale=0.68]{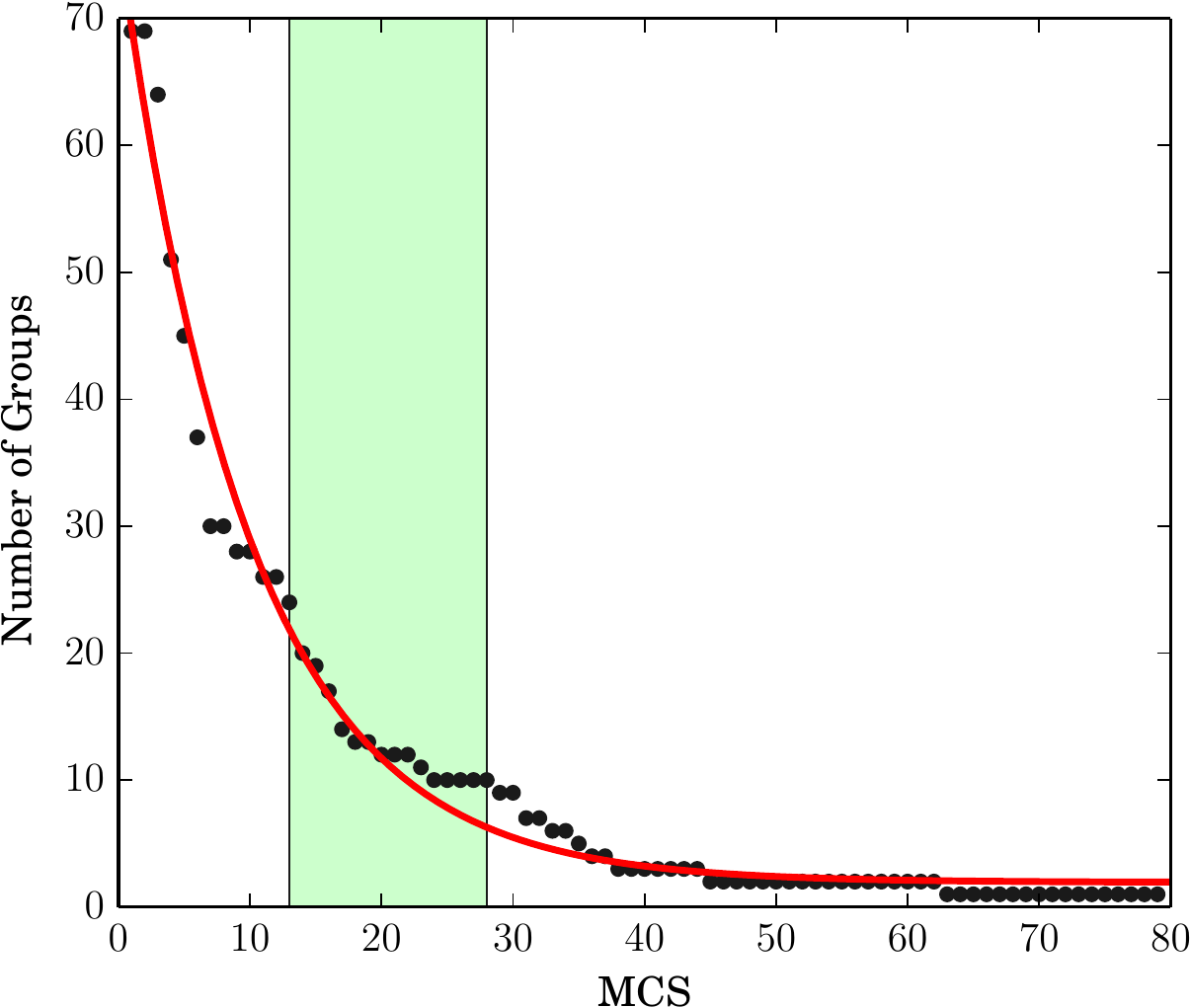}
\caption[Number of resulting circular streams as a function of \textit{Min Cluster Size} parameter]{Number of resulting
circular streams as a function of \textit{minimum cluster size} parameter (MCS).
The red line shows an exponential function fitted to the points. The area of the preferred values for
MCS is marked in green ($\rm 13 < MCS < 28$).}
\label{fig:Cluster_vs_minsize}
\end{center}
\end{figure}

Once the standardization and the uniqueness parameters are fixed, we can study how the number of groups is affected by
the MCS parameter. The MCS parameter is the minimum number of stars that a cluster should have in order not to be
rejected during the process. It has a direct impact on the final number of groups resulting from the study.
Figure~\ref{fig:Cluster_vs_minsize} shows the number of groups found by \beacon~as a function of the MCS parameter.
Optimal values were found in the range of $\rm 13 < MCS < 28$, which are such that the derivative of a fitted
negative exponential $f^\prime({\rm MCS}) = -a\; b \;\exp(-b\; {\rm MCS})$ is between $-1$ and $-0.5$. Below MCS =
13, detection rates increase rapidly with a small decrease in the MCS parameter, indicating possible false detections.
On the other hand, individual group detections drop rapidly to the general rotation pattern of the galaxy for values
greater than MCS = 28, not allowing us to analyze the finer details. A more detailed explanation concerning this choice, and the influence of the MCS parameter on the final results can be found in Sections~\ref{Subcap:Angular_momentum} and \ref{Cap:Consistency}. The \textit{clustering parameters} finally used are
listed in Table~\ref{tab:Clustering_par}.

\begin{table}
\caption{Clustering parameters}
\begin{tabular}{lr}
\hline
\hline
Quantity & Value \\
\hline
MCS &  $[13, 28]$ \\
$\epsilon_{\rm c}$ &  $1/3$ \\
\textit{Standardization Method} & Standard deviation \\
\textit{Uniqueness Criteria} & Any element \\
\hline
\label{tab:Clustering_par}
\end{tabular}
\end{table}

\beacon~has been designed to find chemo-kinematic patterns which differ from the general kinematic behaviour of the stellar system. Therefore, it is unlikely to find large scale kinematics patterns affecting the whole galaxy i.e. a general rotation pattern or entirely random kinematics. This is due to the way clusters are formed in the $\mathbf{\Phi}$ space, where general kinematic patterns involving all stars regardless of their metallicity represent a continuous cloud of points. This is the case of proper--motion induced gradients, such as the one found in Section~\ref{introduccion:magnesium}, which was discarded by \beacon, and not taken into account in the subsequent analysis.

\subsection{Derivation of galaxy parameters using \beacon}
\label{Cap:Par_in Fornax}

Stars in a galaxy in equilibrium should be rotating around the CM of the galaxy. Therefore, \beacon~is expected to find
more BSSs and less OSSs around this CM. Using this idea, \beacon~can be
applied to find the best \textit{galaxy parameters}, i.e. the CM coordinates. This is independent of
the chosen \textit{clustering parameters} and can be done by maximizing the number of BSSs and
minimizing the OSSs as a function of the \textit{galaxy parameters}.

The procedure is as follows. A grid of coordinates is generated centred on the commonly adopted coordinates for the
galaxy: RA$^{\rm gal}$, Dec$^{\rm gal}$ and ${\rm v}^{\rm gal}_{\rm LOS}$. Each set of coordinates from this grid is used as
input \textit{galaxy parameters} for \beacon. The number of both, BSSs and OSSs, is stored
in a data cube of the same dimension as the coordinates grid. The coordinates of the CM should correspond to the point
of the grid producing the largest number of BSSs and the smallest number of OSSs. We
quantify this requirement defining:
\begin{equation}
	\mu({\rm RA}^{\rm CM}, {\rm Dec}^{\rm CM}, {\rm v}^{\rm CM}_{\rm LOS}) =
	\frac{(\left\vert{\rm BSSs}\right\vert +1 )^2}{(\left\vert{\rm OSSs}\right\vert +1 )^2}
\end{equation}
where $\left\vert{\rm BSSs}\right\vert$ is the number of
BSSs and $\left\vert{\rm OSSs}\right\vert$ is the number of OSSs
found for the coordinates (RA$^{\rm CM}$, Dec$^{\rm CM}$, ${\rm v}^{\rm CM}_{\rm LOS}$).

In the case of Fornax, we used the values listed in Table~\ref{tab:Fornax_Glance} as central values for generating the
grid of coordinates with $150 \times 150 \times 200$ steps in RA, Dec, and \vlos ~respectively. The value of $\mu$
was calculated using \beacon~for each of the $4.5 \times 10^6$ grid coordinate sets. This provided the
probability distribution map of the CM coordinates.

\begin{figure}
\begin{center}
\includegraphics[ scale=0.76]{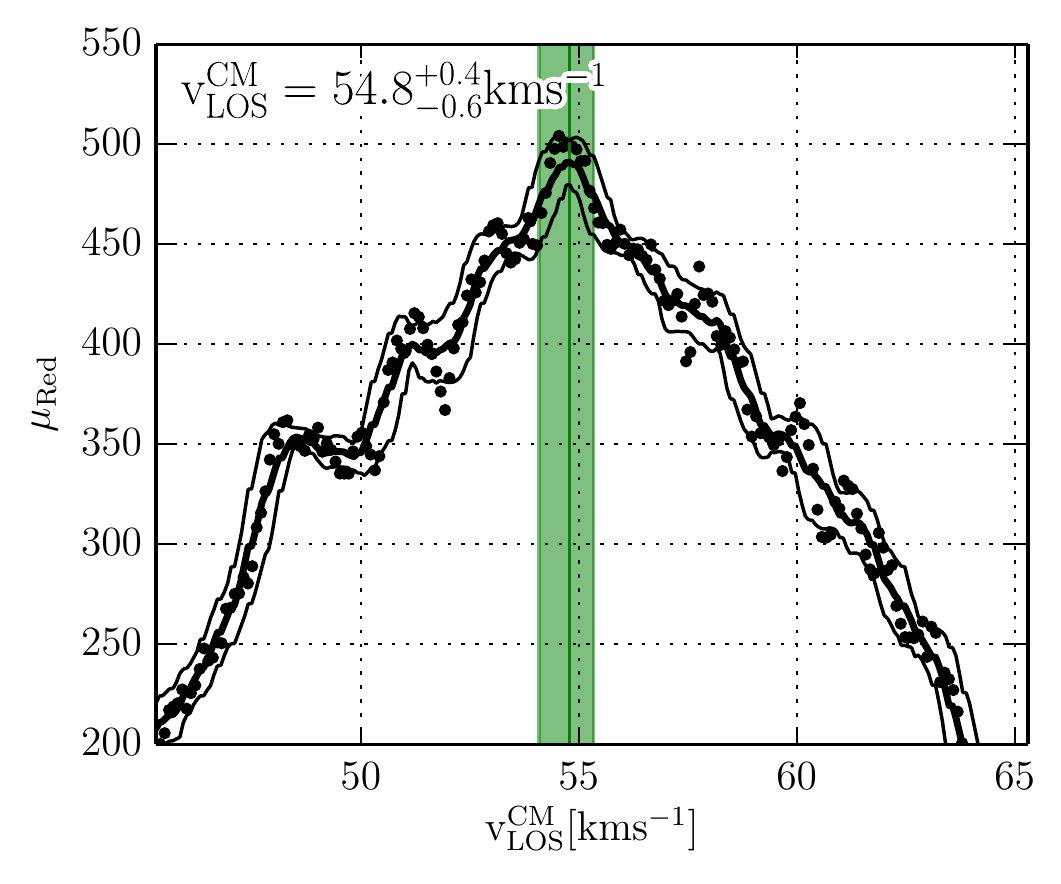}
\caption[\vlos$_{CM}$ of Fornax derived from \beacon]{$\mu_{\rm red}$ as a function of ${\rm v}^{\rm CM}_{\rm LOS}$ for
Fornax. The average value of $\mu_{\rm red}$ and its dispersion are shown by the black thick and thin curves
respectively. The peak of the distribution is marked by the vertical green line while its dispersion is shown with the
shaded area.}
\label{fig:Cluster_vs_vlosCM}
\end{center}
\end{figure}

A two-dimensional Gaussian profile was then fitted to each
$\mu({\rm RA}^{\rm CM}, {\rm Dec}^{\rm CM}, {\rm v}^{\rm CM}_{{\rm LOS}, k})$
distribution map, with $0 \leq k \leq 199$. We define a `reduced' $\mu$, $\mu_{\rm red} ({\rm v}^{\rm CM}_{\rm LOS})$, as
the height of the fitted Gaussian. This definition allows us to find the ${\rm v}^{\rm CM}_{{\rm LOS}, k}$ which produces the
highest density of favourable results, i.e. the maximum of $\left\vert{\rm BSSs}\right\vert$ and the minimum
of $\left\vert{\rm OSSs}\right\vert$. Results of this procedure are shown in Figure~\ref{fig:Cluster_vs_vlosCM}.

\begin{figure}
\begin{center}
\includegraphics[ scale=0.78]{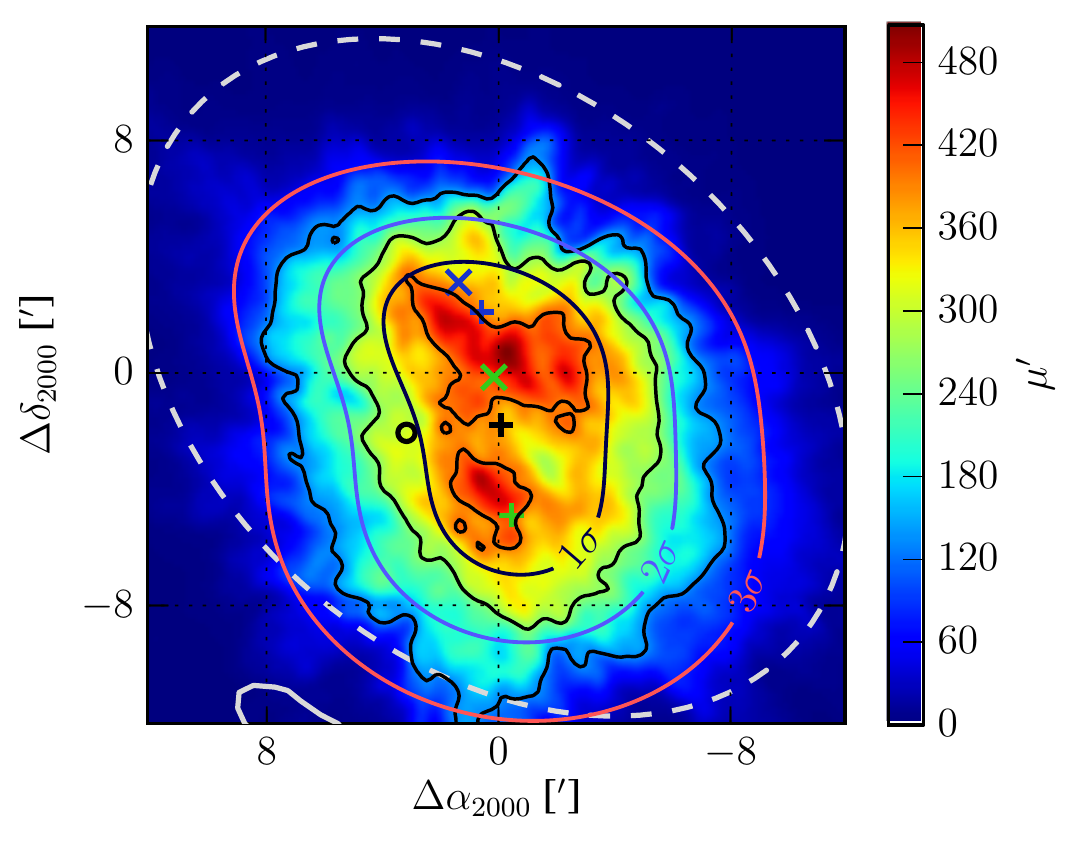}
\caption[The location of the centre of mass of Fornax]{Probability density map of the location of the CM of
Fornax. The color scale represents the value of $\mu^{\prime}({\rm RA^{CM}, Dec^{CM}})$. Contours correspond to the
$3\sigma$, $2\sigma$ and $1\sigma$ differences in standard deviation units with respect to the best solution found
with two Gaussians. The centres of the two-Gaussian fits are represented by the blue and green $+$ signs, respectively
for (1) and (2). The centre obtained from fitting a single Gaussian profile is represented by the black $+$ sign. The
core radius is shown by the light gray dashed ellipse, while the globular cluster Fornax 4 is indicated by the empty
circle. The inner clump of stars found by \citet{Coleman2004} is shown at $\Delta\alpha_{2000} \sim 8^\prime$,
$\Delta\delta_{2000} \sim -10^\prime$. Two optical centres have been plotted with the $\times$ signs: the one by
\citet{Mateo1998} in blue and the one derived by \citep{van_den_Bergh1999} in green.}
\label{fig:CM_MAPS_Data}
\end{center}
\end{figure}

We applied a box-car average with a bin width of 0.6 km s$^{-1}$ (the typical error in \vlos ~of our sample) in 256
steps over $\mu_{\rm red} ({\rm v}^{\rm CM}_{\rm LOS})$. This resulted in a value ${\rm v}^{\rm CM}_{\rm LOS} = 54.8^{+0.4}_{-0.6}
$ km s$^{-1}$ for Fornax, which is in good agreement with results of other authors \citep{Mateo1991,
Walker2006}. We then constructed $\mu^{\prime}(\rm RA^{CM}, Dec^{CM})$ by integrating $\mu({\rm RA^{CM}, Dec^{CM}},
{\rm v}^{\rm CM}_{\rm LOS})$ in the ${\rm v}^{\rm CM}_{\rm LOS}$ direction:
\begin{equation}
	\mu^{\prime}({\rm RA^{CM}, Dec^{CM}}) =
	\int_{v_1}^{v_2} \mu({{\rm RA^{CM}, Dec^{CM}}, {\rm v}^{\rm CM}_{\rm LOS}}) {\rm d} {\rm v}^{\rm CM}_{\rm LOS}
\end{equation}
where $v_1 = {\rm v}^{\rm CM}_{\rm LOS} - \sigma ({\rm v}^{\rm CM}_{\rm LOS})$ and $v_2 = {\rm v}^{\rm CM}_{\rm LOS} + \sigma (
{\rm v}^{\rm CM}_{\rm LOS})$. The resulting map is shown in Figure~\ref{fig:CM_MAPS_Data}. Two possible centres of rotation
were found near the optical centre of the galaxy. The most important one is located roughly at the optical centre
given by \citet{van_den_Bergh1999}. The other centre was identified at $\sim 4^\prime$ to the south of the optical
centre. This may indicate that the galaxy has recently accreted some material which is still not in equilibrium.

We have determined the best \textit{galaxy parameters} for Fornax by fitting $\mu^{\prime}({\rm RA^{CM}, Dec^{CM}})$
with a single Gaussian profile and a profile composed of two Gaussians. The results are listed in
Table~\ref{tab:Galaxy_par_tunned}. The two-Gaussian profile clearly fits the data better. This profile can be used
to obtain $\mu_{\rm red}$ resulting in two preferred velocities for each CM. The one coinciding with the
optical centre of the galaxy has a velocity of ${\rm v}^{\rm CM}_{{\rm LOS},1} = 55.5^{+0.5}_{-0.8}$ km s$^{-1}$, while the
secondary centre shows a systemic \vlos ~of ${\rm v}^{\rm CM}_{{\rm LOS},2} = 54.5 ^{+0.7}_{-0.6}$ km s$^{-1}$. This difference is not significant (within $1\sigma$), although this refers to only one of the three velocity components. We do not have information about the transversal components. On the other hand, the fact that the projected positions of both CM are statistically different, may indicate the presence of at least two dynamically decoupled components in the galaxy.

\begin{table*}
\caption{Tuned galaxy parameters for the Fornax dSph.}
\begin{tabular}{lccr}
\hline
\hline
Model & RA$^{\rm CM}$ & Dec$^{\rm CM}$ & ${\rm v}^{\rm CM}_{\rm LOS}$ \\
       & (J2000) & (J2000) & [km s$^{-1}$] \\
\hline
Single-Gaussian  &  2h 39m $(51.93\pm0.02)$s &  $-34^{\circ}$ $31^{\prime}$ $(54.1\pm0.4)^{\prime\prime}$
& $54.8^{+0.4}_{-0.6}$ \\
Two-Gaussian (1) &  2h 39m $(55.1\pm0.1)$s   &  $-34^{\circ}$ $28^{\prime}$ $(1\pm2)^{\prime\prime}$
& $55.5^{+0.5}_{-0.8}$ \\
Two-Gaussian (2) &  2h 39m $(50.14\pm0.05)$s &  $-34^{\circ}$ $35^{\prime}$ $(1\pm3)^{\prime\prime}$
& $54.5^{+0.7}_{-0.6}$ \\
\hline
\label{tab:Galaxy_par_tunned}
\end{tabular}
\end{table*}

\subsection{The age of the stars}
\label{Cap:Ages}

The existence of an AMR in Fornax has been demonstrated photometrically and spectroscopically in several works
\citep[see for example][]{Pont2004, Battaglia2006, Coleman2008, deBoer2012, delPino2013,
Piatti2014}. In principle, this allows us to assign an average age to each group of stars based on their metallicity.
Several methods exist to derive the ages of spectroscopically measured stars. \citet{Carrera2008a,Carrera2008b} made
use of polynomial relationships in order to obtain the approximate age of each star. In this paper however, we will
use probability distribution functions (PDFs) of the metallicity distribution in each group and fit these to our
observed AMR of Fornax. The reason to do so is the lack of precision in the metallicity measurements for a given star,
which is however not a strong obstacle when considering the metallicity distribution of the whole group. The
procedure is described below.

For each star in a group, we created $2 \times 10^3$ synthetic stars with metallicities chosen randomly from a
Gaussian probability distribution with a mean value equal to the metallicity of the star, $Z_i$, and dispersion
$\sigma$ equal to the metallicity error $\sigma(Z_i)$. In this way, we have a distribution of
$2n\times10^3$ stars for the whole group, where $n$ is the number of stars in the group.

We constructed a PDF for the AMR based on the three SFHs derived in \citet{delPino2013}. We created an average SFH of
the three SFHs normalized to the integral over time of their star formation rate $[\psi(t)]$. From this averaged SFH,
we created a discrete AMR over $5\times10^3$ age steps linearly distributed between 2 Gyr and 13.5 Gyr (we do not expect
RGB stars younger than 2 Gyr). In each one of these age steps, we created a distribution of $10^4$ stars, following the
PDF derived from the SFH at that age. This created a total of $5\times 10^7$ stars, in $5\times10^3$ metallicity
distributions following the metallicity distributions of the SFH at the corresponding age.
The procedure ends by comparing the metallicity distribution for each group with the $5\times10^3$ metallicity
distributions generated from the SFH. The comparison was made through a Kolmogorov-Smirnov (K-S) test and the results
were stored for each group and age step.

In order to avoid stochastic sampling variations in the results, the whole procedure was repeated 1000 times. For each
group, the resulting $p$-values of the K-S test turned out to be normally distributed around a central age value. The
adopted age of each group is therefore the mean value of the normal distribution of all its assigned ages and the
assigned error is the standard deviation of this distribution.

\section{Results}
\label{Cap:Results}

After applying \beacon~over our catalogue of 2562 stars, a total of 985 of them were classified into 24 possible
circular streams. No non-circular streams were detected. The 1577 remaining stars were considered as stars which
do not follow any strong chemo-kinematical pattern. Some of these stars may belong, in fact, to rotating streams and
have been misclassified due to the lack of 3D kinematic information for them. For example, stars with circular orbits 
and projected positions close to their rotation axis will show nearly zero \vlos, making their classification
into BSSs difficult. Some other may simply belong to badly sampled populations that have not fulfilled the minimum requirements to be 
considered as a group. Still, the fact that $\sim 60\%$ of the stars were not classified into BSS indicates that Fornax is 
dominated by entirely random kinematics. This is the expected result for a pressure supported dSph galaxy, and does not rule 
out the presence of rotating stellar populations in the galaxy.

\begin{figure}
\begin{center}
\includegraphics[ scale=0.58]{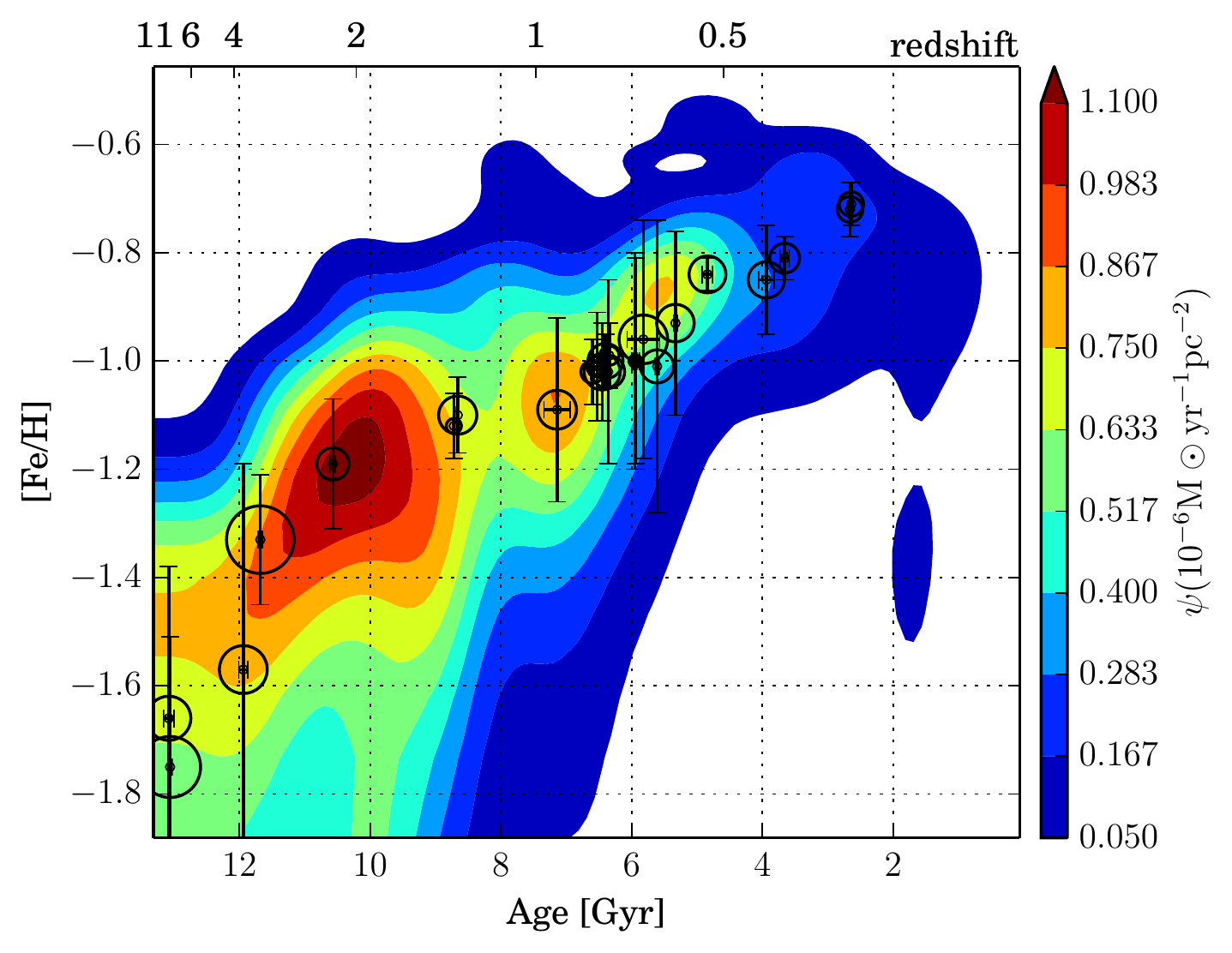}
\caption[The Fornax SFH and the ages of the SC]{Averaged SFH used to derive the ages of the BSSs. Each
BSS is represented by a circle of size proportional to its angular momentum modulus. Vertical error bars
show the metallicity dispersion of each BSS.}
\label{fig:AMR}
\end{center}
\end{figure}

We derived the age for each BSS following the method described in the previous section. The typical uncertainty provided by this method is
of the order of 0.05 Gyr, much smaller than the intrinsic error associated to the SFH of Fornax \citep{delPino2013}, and other possible
sources of error not discussed in this paper. For this reason we do not expect precision higher than $\sim 2$ Gyr at oldest ages, and no
better than $\sim 0.7$ for ages around 3 Gyr. The ages obtained here provide therefore an insight on the sequence of events associated with
the different clusters found by \beacon, allowing us a direct comparison with the SFH of Fornax. The averaged AMR
of Fornax, obtained from the SFH derived in \citet{delPino2013} and the BSS groups with their assigned
ages are shown in Figure~\ref{fig:AMR}.
In addition to the age, a set of other useful properties can be derived for each BSS.
Some of these quantities are listed in Table~\ref{tab:CS_mean_quantities}. In the following we discuss
each of them separately.

\begin{table}
\caption{Mean properties of circular streams detected in Fornax.}
\label{tab:CS_mean_quantities}
\begin{tabular}{@{}lcccr}
\hline
\hline
SC & \feh &  $|\mathbf{L}|$ & $\theta$ & Age\\
             &                & $\rm [10^{3} pc^{2}s^{-1}]$ & $[^\circ]$ &[Gyr]\\
\hline
 1 & $-1.75 \pm 0.24 $ &  $14.0   \pm 0.5$   & $172    \pm 2   $  & $ 13.06 \pm 0.03 $ \\
 2 & $-1.66 \pm 0.28 $ &  $7.3    \pm 0.3$   & $174    \pm 1   $  & $ 13.08 \pm 0.08 $ \\
 3 & $-1.57 \pm 0.38 $ &  $8.6    \pm 0.2$   & $344.5  \pm 0.9 $  & $ 11.94 \pm 0.07 $ \\
 4 & $-1.33 \pm 0.12 $ &  $17.2   \pm 0.7$   & $175    \pm 3   $    & $ 11.68 \pm 0.03 $ \\
 5 & $-1.19 \pm 0.12 $ &  $3.91   \pm 0.08$  & $17.5   \pm 0.7 $    & $ 10.57 \pm 0.04 $ \\
 6 & $-1.12 \pm 0.06 $ &  $0.871  \pm 0.008$ & $358.5  \pm 0.2 $  & $ 8.72  \pm 0.02 $ \\
 7 & $-1.10 \pm 0.07 $ &  $5.5    \pm 0.1$   & $167.9  \pm 0.4 $  & $ 8.66  \pm 0.02 $ \\
 8 & $-1.09 \pm 0.17 $ &  $5.7    \pm 0.3$   & $302    \pm 3   $    & $ 7.14  \pm 0.20 $ \\
 9 & $-1.02 \pm 0.06 $ &  $2.02   \pm 0.07$  & $350.2  \pm 0.6 $    & $ 6.60  \pm 0.03 $ \\
10 & $-1.01 \pm 0.10 $ &  $1.56   \pm 0.06$  & $337.8  \pm 0.6 $  & $ 6.53  \pm 0.04 $ \\
11 & $-1.02 \pm 0.09 $ &  $4.9    \pm 0.2$   & $172    \pm 1   $  & $ 6.45  \pm 0.03 $ \\
12 & $-1.00 \pm 0.05 $ &  $4.6    \pm 0.2$   & $9.7    \pm 0.9 $  & $ 6.40  \pm 0.02 $ \\
13 & $-0.99 \pm 0.06 $ &  $1.4    \pm 0.2$   & $98     \pm 1   $    & $ 6.34  \pm 0.02 $ \\
14 & $-1.02 \pm 0.17 $ &  $4.1    \pm 0.3$   & $48     \pm 2   $    & $ 6.36  \pm 0.04 $ \\
15 & $-1.00 \pm 0.19 $ &  $0.60   \pm 0.01$  & $13.6   \pm 0.3 $  & $ 5.94  \pm 0.06 $ \\
16 & $-1.00 \pm 0.20 $ &  $0.52   \pm 0.01$  & $19.0   \pm 0.3 $  & $ 5.95  \pm 0.03 $ \\
17 & $-1.01 \pm 0.27 $ &  $4.1    \pm 0.2$   & $218    \pm 1   $    & $ 5.61  \pm 0.03 $ \\
18 & $-0.93 \pm 0.17 $ &  $5.3    \pm 0.3$   & $7.3    \pm 2   $  & $ 5.33  \pm 0.03 $ \\
19 & $-0.84 \pm 0.03 $ &  $5.1    \pm 0.1$   & $80.7   \pm 0.6 $  & $ 4.84  \pm 0.08 $ \\
20 & $-0.96 \pm 0.22 $ &  $8.9    \pm 0.3$   & $25     \pm 3   $  & $ 5.8   \pm 0.2  $ \\
21 & $-0.85 \pm 0.10 $ &  $5.0    \pm 0.2$   & $136    \pm 1   $  & $ 3.9   \pm 0.1  $ \\
22 & $-0.81 \pm 0.04 $ &  $3.3    \pm 0.1$   & $177.5  \pm 0.5 $  & $ 3.66  \pm 0.07 $ \\
23 & $-0.72 \pm 0.05 $ &  $2.5    \pm 0.2$   & $133.4  \pm 0.7 $  & $ 2.66  \pm 0.03 $ \\
24 & $-0.71 \pm 0.04 $ &  $2.2    \pm 0.1$   & $343.8  \pm 0.5 $  & $ 2.64  \pm 0.03 $ \\
\hline
\end{tabular}
\end{table}

\subsection{The angular momentum}
\label{Subcap:Angular_momentum}

Probably the most interesting property which can be derived  for each BSS is the projected angular
momentum, $\mathbf{L}  = \sum_{i=1}^{i=j}{\mathbf{r}_i \times m_i\mathbf{v}_i}$, where $m_i$ is the mass of the
$i^{\rm th}$ star and $j$ is the number of stars in the BSS.
A priori, we do not have information about the homogeneity of our sample in terms of metallicity. A sensible way to
avoid sampling effects is to normalize the $\mathbf{L}$ vector of every group to the number of stars in the group.
This is roughly the same as expressing the angular momentum per unit of stellar mass and provides a measure of the mean
momentum $\mathbf{p}$ of the stars forming a BSS. In the following we will use only
quantities normalized in this way to the number of stars in each BSS.

\begin{figure}
\begin{center}
\includegraphics[ scale=0.6]{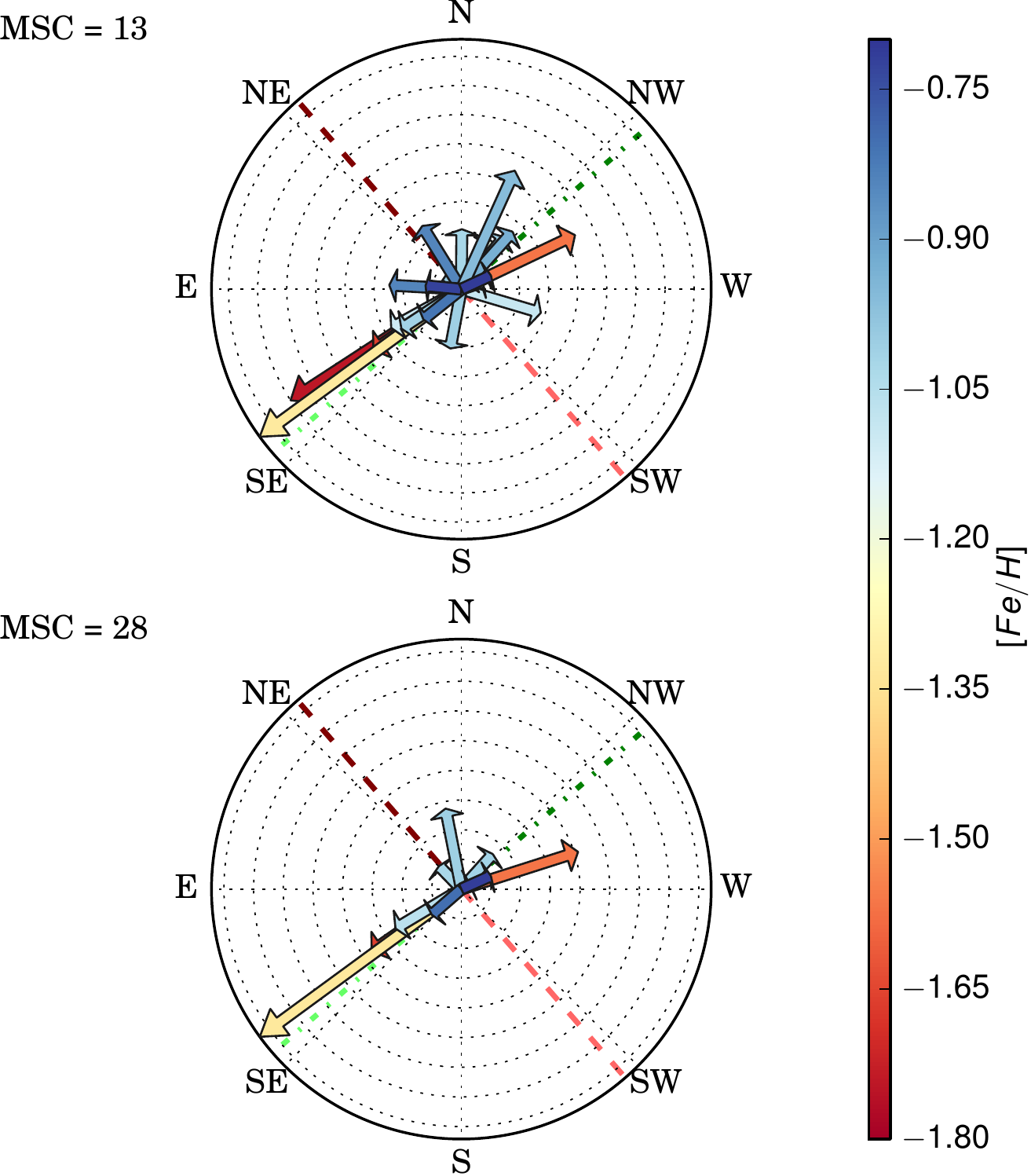}
\caption[Angular momenta of the different rotating stellar populations in Fornax.]{Angular momenta of the different
rotating stellar populations in Fornax. Top: the angular momenta $\mathbf{L}$ of the 24 circular
streams BSS identified with MCS = 13. Colours represent the mean metallicity of the group. Bottom: same
as the top panel, but using MCS = 28. Green dotted-dashed lines represent the orientation of the optical minor axis of the galaxy. The optical major axis is shown by the red dashed lines. Darker colours indicate the positive direction along the axes.}
\label{fig:Angular_momentum}
\end{center}
\end{figure}

\begin{figure}
\begin{center}
\includegraphics[ scale=0.6]{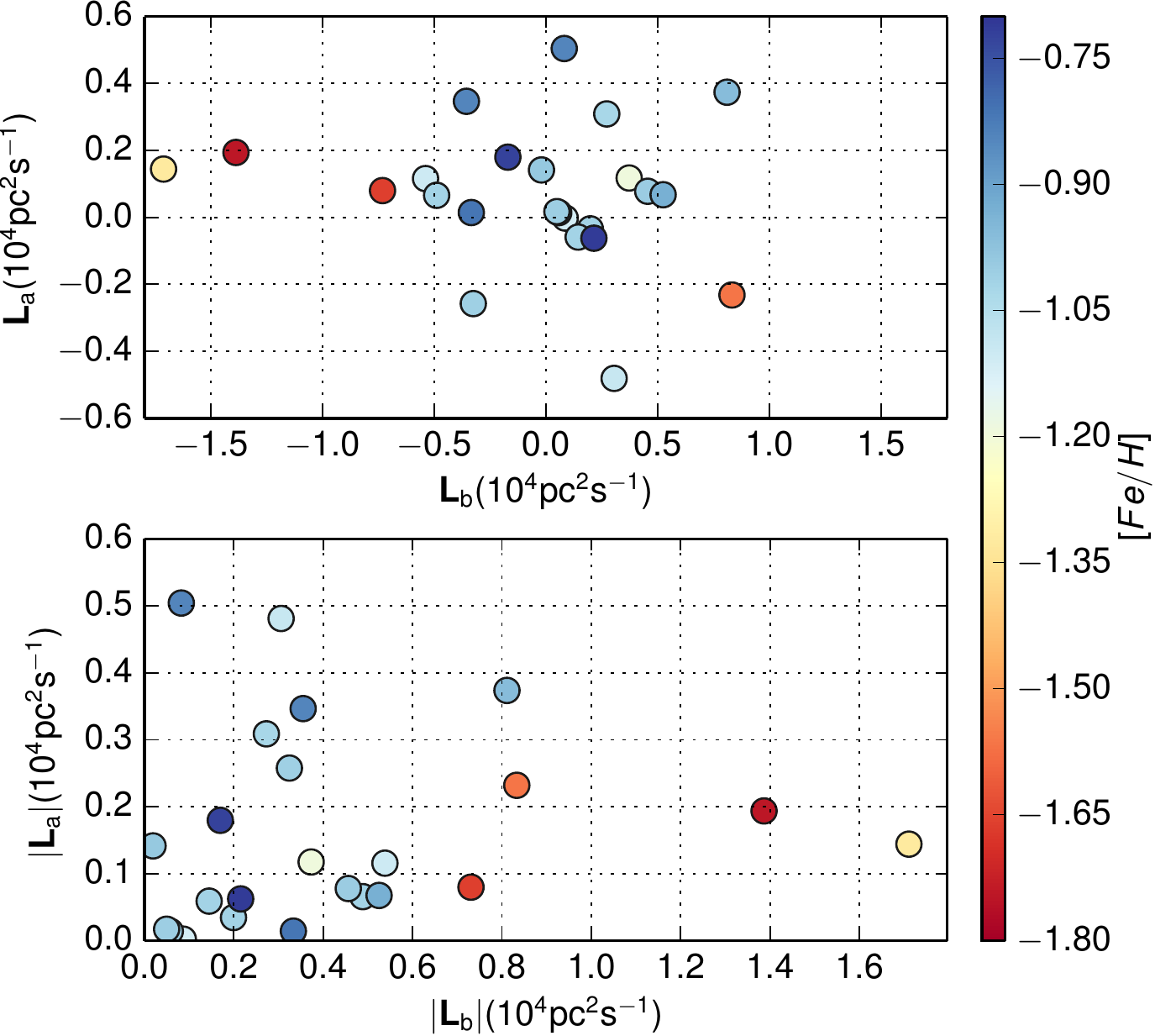}
\caption[Projection over the axes of the system of the angular momenta]{Projection of the angular momenta,
$\mathbf{L}$, on to the major ($a$) and minor ($b$) symmetry axes of Fornax for the 24 BSSs. Colours
represent the mean metallicity of the groups.}
\label{fig:La_Lb}
\end{center}
\end{figure}

Figure~\ref{fig:Angular_momentum} shows the angular momenta $\mathbf{L}$ for the 24 BSSs found using MCS
= 13 and the 10 BSSs found adopting MCS = 28. The resulting $\mathbf{L}$ show a clear preferred
orientation. While the low-metallicity BSSs tend to rotate around the minor axis of the galaxy, the more
metal rich ones appear to have more random patterns of rotation. Low-metallicity BSSs show larger
$|\mathbf{L}|$ in comparison with groups of higher metallicity. The latter can be clearly seen from the
projections of the angular momenta on to the main symmetry axes of the galaxy. We will refer to these as $a$ for the
major axis and $b$ for the minor axis. The projections are shown in Figure~\ref{fig:La_Lb}. Groups of stars with low
\feh~ show larger projections of $\mathbf{L}$ on to the minor axis $b$. This does not mean that metal poor stars
rotate faster, because they are also more spatially extended than more metal rich stars.

The alignment of the strongest $\mathbf{L}$ with the main projected optical axes of Fornax is the expected result from
the clustering of stars in a rotation-supported system. Interestingly, we observe a small misalignment ($\sim
7^{\circ}$) between the observed optical minor axis of Fornax and its dominant angular momentum vector. This may
indicate that the galaxy is in reality a projected triaxial system, rather than an oblate one \citep{Binney1985,
Weijmans2014}. This is also consistent with the shape expected from $N$-body simulations of initially disky dwarf
galaxies subject to tidal interactions with the Milky Way. As discussed in
detail by \citet{Lokas2014a, Lokas2015}, the stellar component of such a dwarf galaxy typically transforms into a
triaxial (but mostly prolate) bar-like system at the first pericentre passage. This shape becomes more spherical during
the subsequent evolution but remains prolate and retains some rotation around the minor axis.

\begin{figure}
\begin{center}
\includegraphics[ scale=0.6]{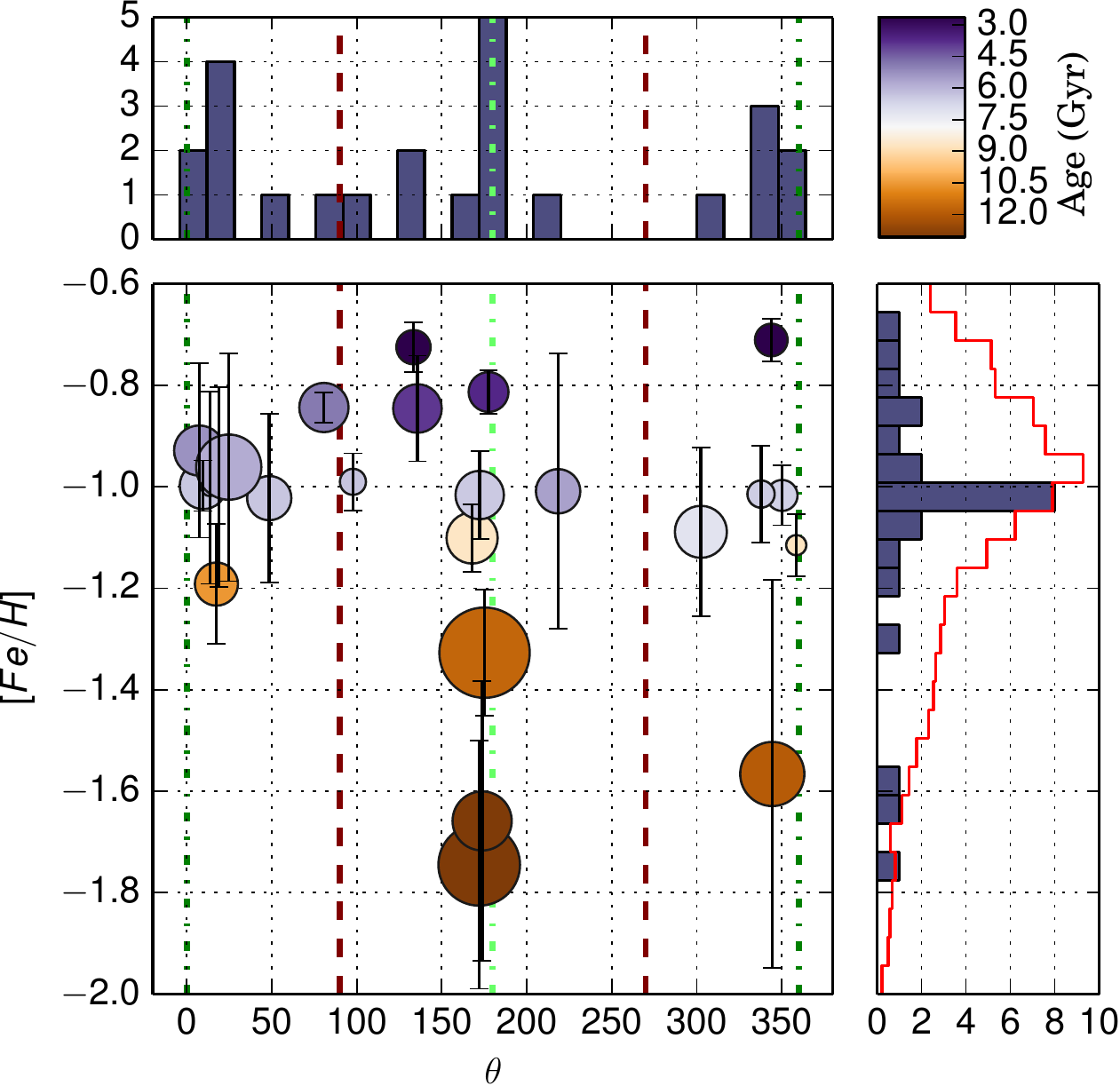}
\caption[Metallicity vs. orientation]{The metallicity, \feh, of the BSSs as a function of the
orientation, $\theta$, of their angular momenta, $\mathbf{L}$. Circles represent the BSSs. Their average
ages are indicated with a colour scale while the radii of the circles are proportional to $|\mathbf{L}|$. Error bars
represent the metallicity dispersion of the groups. Main axes of Fornax are shown by vertical lines. As in Figure~\ref{fig:Angular_momentum}, the
dotted-dashed green lines represent the optical minor axis, while the dashed red ones the major axis. Angles
are measured from the positive semi-minor axis of the galaxy, $+b$. Along both axes of the
figure we plot histograms showing the distribution of the BSSs as a function of
$\theta$ and \feh. The metallicity distribution of all stars in our sample is drawn with a red line.}
\label{fig:Z_A}
\end{center}
\end{figure}

High-metallicity BSSs gather around relatively low values of $L_a$ and $L_b$, some of
them showing a larger projection on to the major axis of Fornax (prolate rotation). We have defined the positive
semi-major and semi-minor axis of Fornax, $+a$ and $+b$ respectively, as the closest to the north. With this
definition, we can express $\mathbf{L}$ as the absolute value of the angular momentum and the angle $\theta$ with
respect to $+b$. In Figure~\ref{fig:Z_A} we show the metallicity of each BSS as a function of
$\theta$. Different behaviours can be observed as a function of metallicity (\feh) and age of the groups. Groups with
\feh~ $\lesssim 1.2$ are older than $\sim10$ Gyr and are arranged along $\theta \sim 350^\circ$ and $\theta \sim
170^\circ$ showing small deviations from the minor axis. The BSSs with $-1.2 \lesssim$ \feh~$\lesssim
-1.0$ exhibit ages from 10 to 6 Gyr. The $\mathbf{L}$s of these groups are more randomly distributed,
some of them showing rotation around the north and south direction. The bulk of BSSs we found gather
around these metallicities, peaking at \feh~$\sim -1.05$. Interestingly, the peak of the metallicity distribution for all
the stars in the sample is located at slightly higher metallicities, \feh~$\sim -0.95$. In principle, one would expect
a peak of detections around the peak of the metallicity distribution of the galaxy. This result indicates that stars
with metallicities \feh~$\sim -1.05$ show a more singular dynamical pattern. The younger BSSs appear to
start to recover the main rotation direction around the minor axis. For the whole galaxy, the preferred
orientations of $\mathbf{L}$ are the main axes of the galaxy, especially the minor axis.

We further tested the effects of changing the MCS parameter from 13 to 28. As it can be seen in
Figure~\ref{fig:Angular_momentum}, some groups disappeared or were merged with bigger ones when using MCS = 28. The groups most susceptible to disappear
were those with metallicities \feh~$\sim -1.05$. This is caused by the more erratic kinematics of the stars at these metallicities.
Interestingly, most of the stars in Fornax have metallicities around these values. This discards any possible statistical problem related
to the number of stars in our sample around these metallicities. The fact that these groups disappear or rearrange when increasing MCS
indicates that stars with \feh~ $\sim -1.05$ possess, indeed, more chaotic kinematics. This appears to be the reason why it is harder for
\beacon~to form groups around these metallicity values when using large MCS requirements. On the other hand, reducing MCS will allow \beacon~ to 
find numbers of groups showing large variety of kinematics. The three main metal--poor components were always detected by \beacon~
for values of MCS below 45. These groups represent the most important contribution to the total angular
momentum of the galaxy. This result was expected since old, metal--poor stars constitute the majority of stars in Fornax \citep{delPino2013}.
Results for high metallicity groups (\feh~ $> -0.8$) were also stable and almost independent of the adopted MCS value.
All these results are consistent with the odd kinematics shown by the stars with metallicities \feh~ $\sim -1.05$.

\subsection{Evolution with time}

Estimating the ages of the BSSs allows us to derive the rotation history (RH), $\mathbf{L}(t)$, of
Fornax. Considering that time is the most important variable in the problem, we define the RH as an explicit function
of it. Formally, $\bar{\mathbf{L}}(t){\rm d}t$ is the predominant angular momentum of the galaxy within the time
interval $(t,t+{\rm d}t)$, which can be calculated as the vector sum of all the $\mathbf{L}_i$ of groups with
ages within that time interval. To give an estimate of how dispersed the $\mathbf{L}_i$
are from the average rotation axis we defined the dispersion around the mean axis as:
\begin{equation}
	\sigma[\bar{\theta}(t)]_{\rm axis} =  \left(
	\frac{\sum_{i=1}^{i=j}{|\mathbf{L}_i|\zeta_i}}{\sum_{i=1}^{i=j}{|\mathbf{L}_i|}}\right)^{1/2}
\end{equation}
where
\begin{equation}
                 \phi_i = \left\{\begin{array}{ccc}
                 \theta_i,                  & {\rm if} & |\zeta_i|  >   \pi/2 \\
                 \theta_i + \pi \pmod 2\pi, & {\rm if} & |\zeta_i| \leq \pi/2 \\
                 \end{array}\right.
\end{equation}
and
	\begin{equation}\zeta_i = \arctan\{\sin[\phi_i-\bar{\theta}(t)], \cos[\phi_i-\bar{\theta}(t)]\}.
\end{equation}
Therefore $\sigma[\bar{\theta}(t)]_{\rm axis}$ represents the dispersion of the $\mathbf{L}_i$ within $(t,t+{\rm d}t)$
around the main axis of rotation $\bar{\mathbf{L}}(t){\rm d}t$, independently of the rotation direction.

\begin{figure}
\begin{center}
\includegraphics[ scale=0.6]{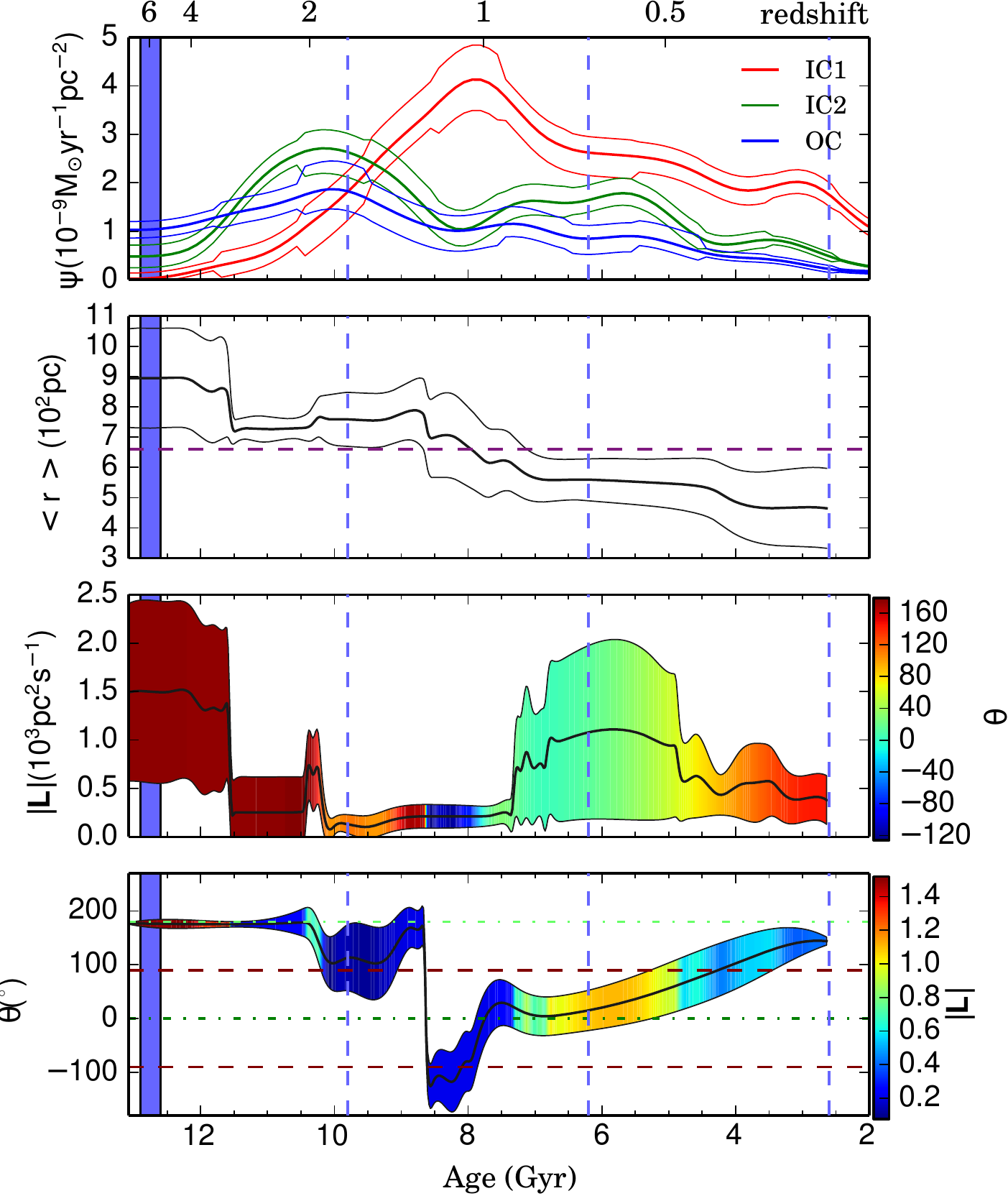}
\caption[Evolution of preferred angular momentum with time (RH)]{Evolution of preferred angular momentum with time.
The first panel shows the SFH of Fornax, derived in \citet{delPino2013} for the
three defined regions (IC1, IC2, OC). The second panel shows the evolution of the mean distance of the stars from the
centre of the galaxy, with the present half-light radius ($r_h$) of Fornax indicated by a horizontal purple dashed
line. The third panel shows the evolution of $|\mathbf{L}|$ with time with the colour scale representing the
orientation ($\theta$) of $\mathbf{L}$. The last panel shows the evolution of $\theta$ with time with the colour scale
representing $|\mathbf{L}|$. As in Figure~\ref{fig:Angular_momentum} the
dotted-dashed green lines indicate the optical minor axis, while the dashed red ones the major axis. The
dispersion of the results is plotted with thinner lines. Possible passages of Fornax through its perigalacticon,
derived from the orbital parameters by \citet{Piatek2007}, are marked by vertical blue dashed lines.
The blue vertical strip in every panel indicates the epoch of reionization.
}
\label{fig:Age_KLA}
\end{center}
\end{figure}

In order to have statistically reliable results, we used a 1.5 Gyr wide bin, shifting it by 512 small steps within
the range of ages found for the BSS. Figure~\ref{fig:Age_KLA} shows the SFH of Fornax derived in
\citet{delPino2013}, $\psi(t)$ (first panel), the mean distance of the stars to the centre of the galaxy, $r$ (second
panel), $|\mathbf{L}|(t)$ (third panel), and the angle between $\mathbf{L}(t)$ and $+b$ (fourth panel).

Our results show a clear evolution in the rotation patterns of Fornax. Older stars rotate around the $-b$ axis (the
minor axis of the galaxy) with very little dispersion. The value of $\bar{\theta}(t)$ suffers a small variation $\sim
10$ Gyr ago towards the $+a$ semi-major axis, but afterwards goes back to the initial rotation direction. The
most important feature in $\bar{\theta}(t)$ occurs $\sim 8$ Gyr ago when the galaxy appears to change its preferred
rotation direction by $\sim120^\circ$. This sudden change has small significance because of the low $|\mathbf{L}|(t)$
at this period, but most importantly, this small $|\mathbf{L}|(t)$ can be the consequence of the multiple and more
chaotic distribution of $\theta_i$ between $\sim10$ and $\sim8$ Gyr ago. The last $\sim5$ Gyr up to 2 Gyr ago show a steady recovery of
the original rotation direction.

Fornax shows an important correlation between the most important features of its SFH and its RH. Star formation peaks,
located at 10 Gyr ago in the outermost regions (IC2 and OC) and at 8 Gyr ago in the central region (IC1), are
accompanied by important changes in the rotation patterns of the galaxy.

The mean distance of the stars from the CM also changes with age. The oldest stars are uniformly spread, even
beyond the core radius. As we move to younger populations, these tend to be more concentrated in the central regions of
the galaxy. This should be taken into account since the scale length of the population is related to its
$|\mathbf{L}|$ momentum. No correlation of $|{\rm v}_{\rm LOS}|$ with age was found in the BSS stars, and this
indicates that higher $|\mathbf{L}|$ originate from the larger scale length of the old populations.

\subsection{Main rotating components}\label{Subcap:Main_rotation}

\begin{figure*}
\begin{center}
\includegraphics[ scale=0.75]{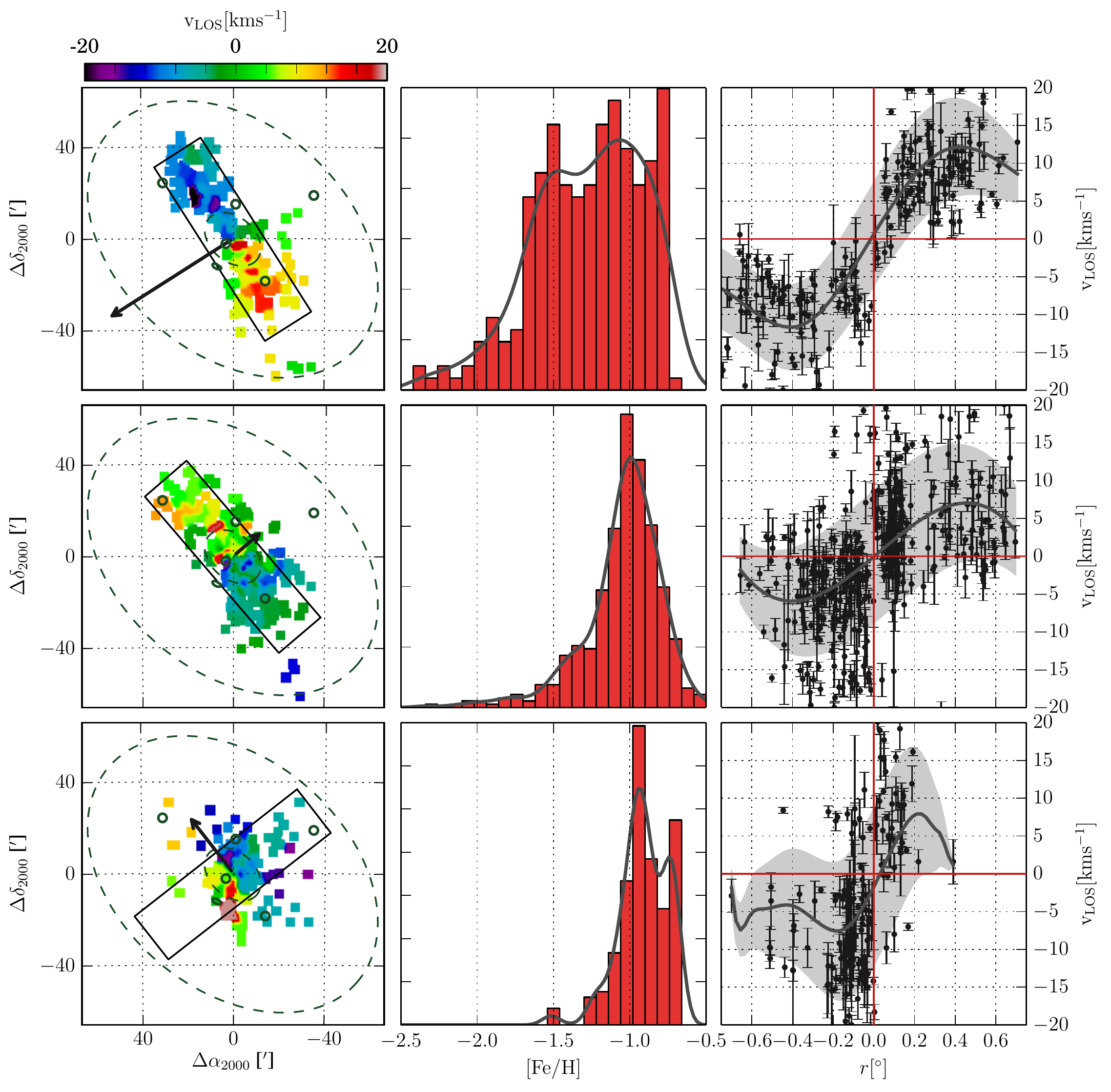}
\caption[Line of sight velocity maps]{Same as Figure~\ref{fig:rawmaps}, for the different BSSs grouped
by their $\mathbf{L}$. Left panels show the \vlos ~maps with the black arrow representing $\mathbf{L}$. Middle panels
show the metallicity distribution of the stars shown in the left panels. Right panels plot the velocity of the stars
along the stripe shown in the left panels. Results from top to bottom are ordered by the average metallicity of the
groups from the lower to higher metallicity. The notation and line types are the same as in Figure~\ref{fig:rawmaps}.}
\label{fig:Maps_KLA}
\end{center}
\end{figure*}

The BSSs can be grouped by their $\mathbf{L}$s. We applied the {\sc Optics} algorithm over the 24
BSSs, clustering them depending on the direction of their $\mathbf{L}$. Three main groups were
found, each rotating around a different main axis of the galaxy. The results are illustrated in
Figure~\ref{fig:Maps_KLA}.

Interestingly, a different metallicity distribution is found for each preferred direction. The main rotation signal
is around the minor axis of the system.
The second group of BSSs rotates in the opposite direction, but with larger velocity
dispersion. Its metallicity distribution peaks at \feh$\sim-1.1$, with a tail towards lower metallicities, indicating
that these stars are at least $\sim8$ Gyr old, some of them reaching the oldest ages found in Fornax. The last group shows
rotation around the optical major axis of the galaxy (prolate rotation). The metallicity distribution of this group
peaks at \feh$\sim-0.9$ and at \feh$\sim-0.7$, which roughly corresponds to stars born $\sim 6$ and $\sim 2$ Gyr
ago, respectively.

\subsection{Rotation curve, orientation and enclosed mass}\label{Subcap:Mass}

\begin{figure}
\begin{center}
\includegraphics[ scale=0.7]{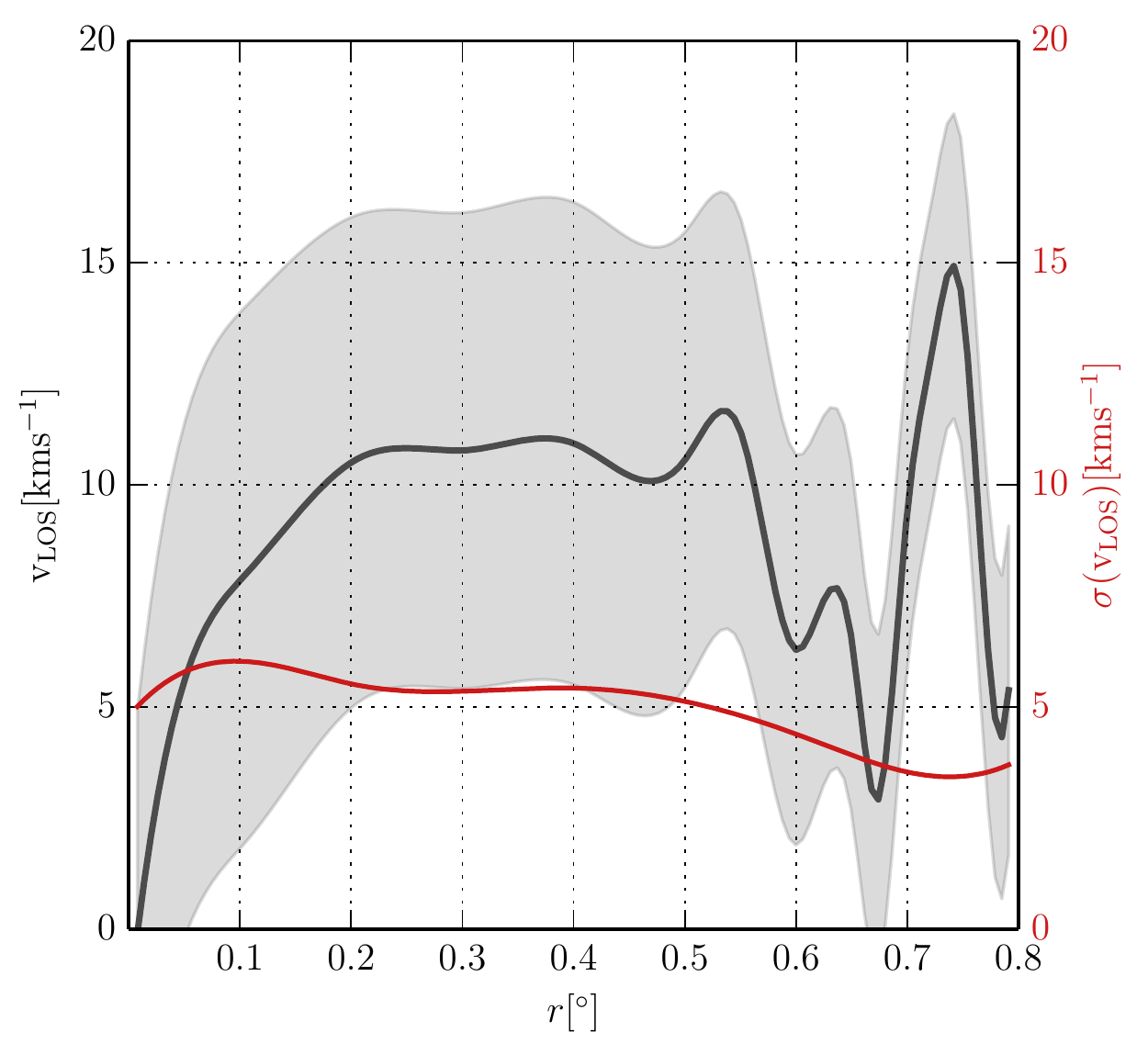}
\caption[Line of sight velocity curve]{Line-of-sight velocity (\vlos) curve (grey line) for the stars along the stripe
shown in the left panel of Figure~\ref{fig:Maps_KLA} for the metal poorer population (first one). The spatial
coordinate has been inverted, as well as the velocity of the stars in the negative axis of the stripe. The velocity
dispersion of these stars is represented by the dashed area and its absolute value by the red curve.}
\label{fig:Rotation_Curve}
\end{center}
\end{figure}

In Figure~\ref{fig:Rotation_Curve} we show the rotation curve extracted from the first population of stars in
Figure~\ref{fig:Maps_KLA}. We have selected this group because it is the one which shows the strongest rotation signal.

The detection of rotating components allows, in principle, to derive the enclosed mass as a function of radius. Assuming spherical distribution, the enclosed mass within a radius, $r$, would be 
\begin{equation}
	M_{dyn}(<r) = r ({\rm v_{c}}^2+\sigma^2) / G
\end{equation}
where $G$ is the gravitational constant, $\rm v_{c}$ the rotation velocity, and $\sigma$ the velocity dispersion. At a first approximation, we can associate the \vlos~of our clusters to the rotation velocity of the galaxy in order to obtain its mass profile. Using values from Figure~\ref{fig:Rotation_Curve}, we obtained $M(r_{half}) = 2.5^{+2.5}_{-1.5}\times 10 ^ 7 M_{\odot}$, assuming $r_{half}=710 \pm 77$ pc \citep{McConnachie2012}. This value is slightly smaller than those obtained through more classical analyses \citep{Lokas2009, Walker2009b, Wolf2010, Walker2011}, but still consistent within $1.5\sigma$ with the average of these. On the other hand, we expect our mass to be underestimated due to projection effects. In fact, our measured $\rm v_{\rm LOS}$ would be the projection of $\rm v_{c}$ along the line--of--sight. Therefore, $\rm v_{c} = \rm v_{\rm LOS}/\cos(\omega)$, where $\omega$ is the angle between the rotation plane and the line--of--sight. Assuming an average of these previous results as true value for $M(r_{half})$, we would need an angle $\omega \sim 50$ deg to reproduce this mass.

\section{Consistency of the results}\label{Cap:Consistency}

Our results are subject to a number of observational effects that may bias them.
The most important one concerns the spatial coverage
of the stars in the sample. If the stars are not uniformly sampled and do not cover the whole body of the galaxy,
this can predispose the experiment to find stars with privileged coordinates. In our case, this can be a sensitive
matter specially for stars lying outside the galaxy core (see Figure~\ref{fig:datasets}), where a clear bias towards
the major axis of the galaxy exists.

\begin{figure}
\begin{center}
\includegraphics[ scale=0.65]{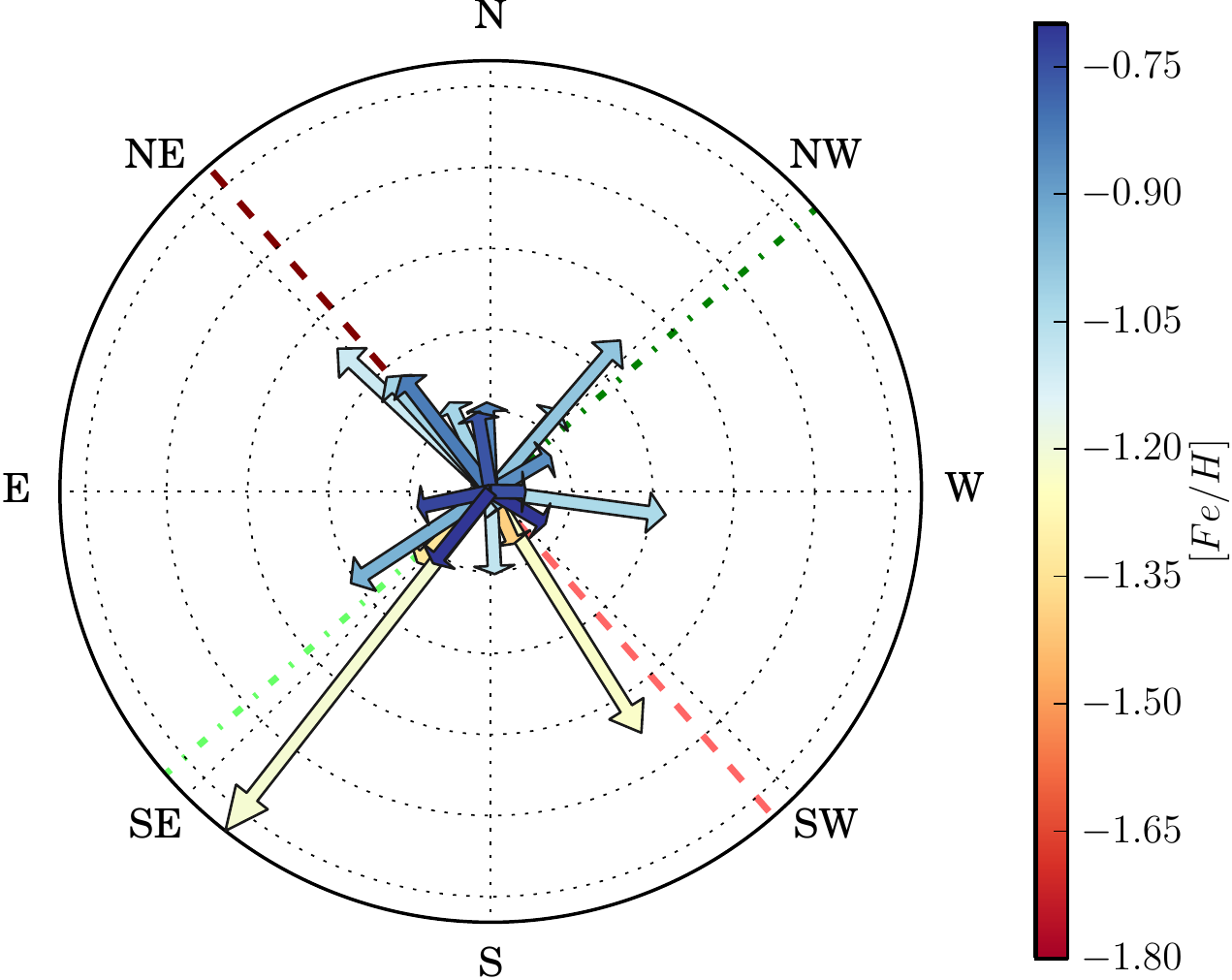}
\caption[Angular momenta in the central region of Fornax.]{Same as Figure~\ref{fig:Angular_momentum} but
for stars within a circle of $7.7^\prime$ in radius (see Figure~\ref{fig:datasets}). The arrows show the angular
momentum vectors, $\mathbf{L}$, while the colours represent the mean metallicity of the group.}
\label{fig:Angular_momentum_centre}
\end{center}
\end{figure}

In order to quantify whether our results can be concentrated around the main axes of the galaxy because of the biased
spatial coverage of our sample, we performed the same experiment using only stars located at the very centre of the
galaxy, within a circle or radius $r=7.7^\prime \sim 300$ pc (see Figure~\ref{fig:datasets}). This assures a relative
isotropy of the sample used in both spatial coordinates $(r,\theta)$. This selection includes approximately $1/3$ of
the original sample, making it necessary to change the MCS parameter to 5. Despite the reduced number of
stars, \beacon~still finds groups with rotation preferably around the main axes of the system (see
Figure~\ref{fig:Angular_momentum_centre}). Since at this radius our data are not spatially biased, the fact that the
strongest angular momenta are aligned with the main optical axes of the galaxy strongly supports the conclusion that
stars indeed move faster around these axes.

A galaxy may be described, at least at some stages of its evolution, as a system of self-gravitating fluid in which
geometry and dynamics go hand in hand. If the system has a preferred angular momentum, $\mathbf{L}$, it will flatten in
the direction perpendicular to $\mathbf{L}$. In fact, assuming a hydrodynamical model for a self-gravitating fluid, one
could infer the shape of the galaxy, once its $\mathbf{L}$ is known. The latter dependence supports our conclusion that
Fornax is rotating mainly around its main optical axes and that this result is not biased by the spatial coverage of our
sample.

\begin{figure}
\begin{center}
\includegraphics[ scale=0.7]{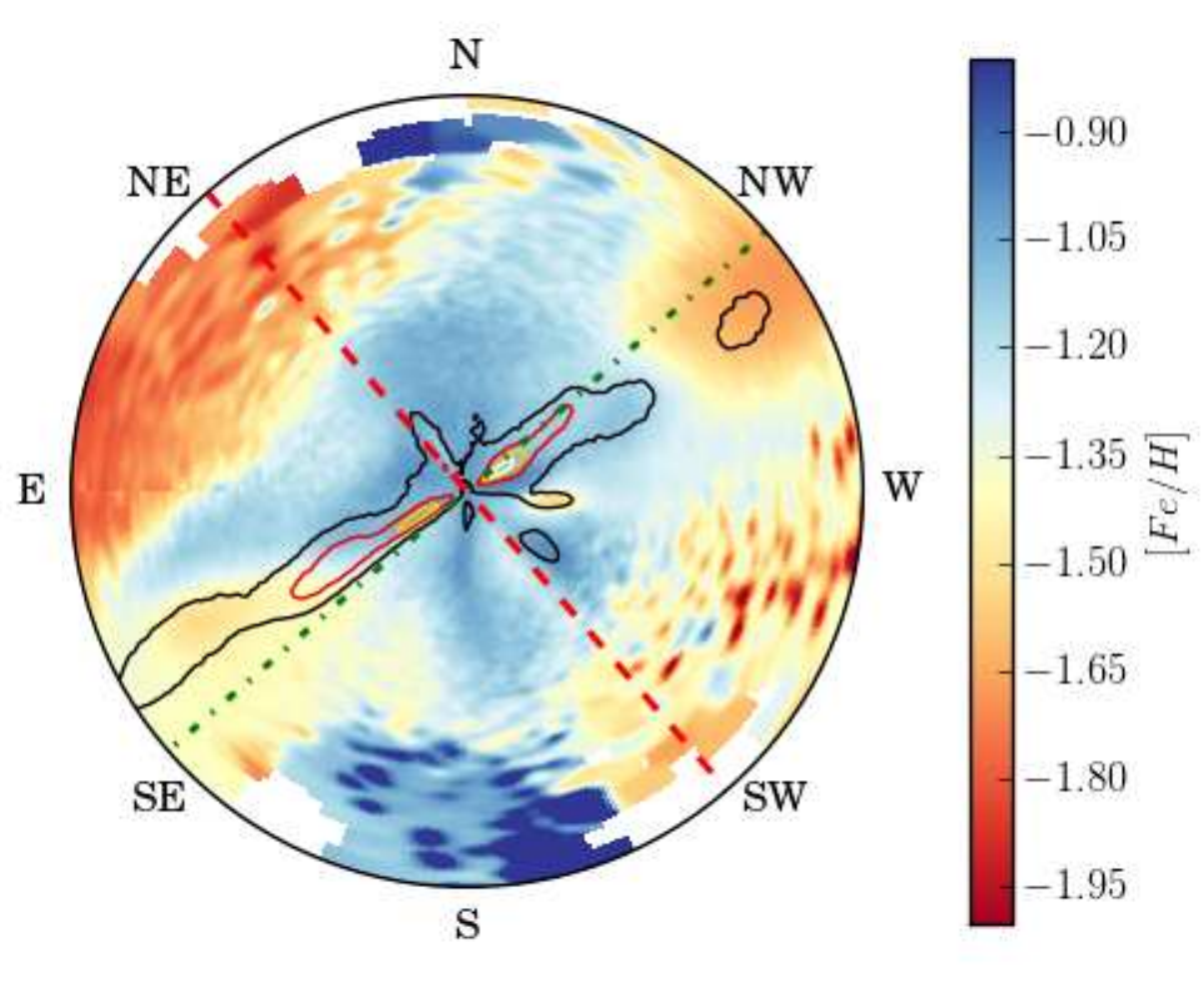}
\caption[Angular momenta obtained from the Monte-Carlo errors simulations.]{The
angular momentum, $\mathbf{L}$, of Fornax calculated with Monte Carlo error simulations over \vlos ~and \feh ~variables.
Colours represent the average metallicity while $1\sigma$, $2\sigma$ and $3\sigma$ contours show the density of the
solutions.}
\label{fig:MC_tests}
\end{center} \end{figure}

The other important source of error in the results are the observational uncertainties of the variables implied
in the clustering process. Assuming no errors in the position of the stars, these uncertainties are $\sigma$(\feh) and
$\sigma$(\vlos). The effects of these in the final results were calculated from Monte Carlo realizations by
stochastically varying the input parameters \feh ~and \vlos ~according to Gaussian probability distributions with
$\sigma$ similar to the observational uncertainties of these quantities. The experiment was repeated $10^6$ times,
obtaining the corresponding BSSs each time. Figure~\ref{fig:MC_tests} shows the results from the Monte
Carlo tests which demonstrate great stability of the results against perturbations in the input variables. Within
$3\sigma$, the rotation of Fornax is almost exclusively around the minor axis $-b$, with some weaker rotation around
the $+b$ axis. This allows us to conclude that our sample is not biased and that our results are stable.

To check the consistency of results as a function of the \textit{clustering parameters}, we ran the code adopting
different values for MCS. This parameter has the biggest impact on the final results, forcing the code to cluster stars
more distant in the space defined by $\mathbf{\Theta}$. We tried all possible values between the maximum and minimum
optimal values (13 and 28). Results were consistent and qualitatively equivalent, producing less groups as we incremented
the MCS parameter, but conserving the general trends in the angular momentum orientation and strength. For example, using
MCS = 28 we obtained 10 BSSs which were very similar to the most important ones found with MCS = 13
(see Figure~\ref{fig:Angular_momentum}). We carefully checked these results using two tests.
In the first one, we crossmatched the stars grouped using MCS = 13 with those using MCS = 28. Only 52
stars were present in the results using MCS = 13 that were not in the ones obtained with MCS = 28. In the latter groups,
we found 23 stars that were not common with the stars in groups using MCS = 13. This represents around 7.6\% of different
stars between both results. As a second test, we divided both results into metal--rich components ([Fe/H] $>$ 1.2) and
metal--poor ([Fe/H] $<$ 1.2). We compared the average angular momenta and metallicities of these components for the results
using both MCS values. No statistically significant differences were found between the results obtained using MCS 28 or 13. On the other
hand, pushing MCS to lower values than 13 produced more stochastic and noisy results. For example, the variations in the clustered
stars for the smallest groups (10 stars or less) using MCS = 9 or 10 were found to be around 15\%.

Lastly, $N$-body simulations were used as input for \beacon~in order to test how well the software can recover different
stellar populations. For this purpose we used the final output of the simulation from \citet{Lokas2014b} aimed to
reproduce the Andromeda II dSph galaxy via a merger of two initially disky dwarf galaxies \citep[see
also][]{EbrovaLokas15}. The $4\times 10^5$ stars of the simulated galaxy were divided in
two different stellar populations with different rotation patterns and metallicities. Specifically, we constructed mock
normal metallicity distributions for the two rotating stellar components, one centred at \feh ~$= -1.3$ and the other
at \feh ~$= -0.9$, with $\sigma = 0.5$ and $\sigma = 0.25$. Several lines of sight were used to create an observational
vector $\mathbf{\Phi}$ of a random selection of 2500 stars containing the projections of the stellar positions on to
the sky, the velocities \vlos ~and metallicities. To make the results more realistic, we simulated artificial
errors in \vlos ~and metallicity using the procedure described in \citet{delPino2015}. For each star, a random value
was extracted from a normal distribution centred on the average error of each variable, with the corresponding
dispersion.

\beacon~succeeded in recovering both populations in a large variety of mock observation set-ups.
In the most favourable
case of the galaxy observed along its intermediate axis, approximately $\sim$75\% of the stars are assigned to
streams. Of these, only 1.2\% were allocated to the wrong group.
In the least favourable case, the galaxy would be observed along the
axis of maximum rotation, which for Andromeda II is the longest axis. In this scenario, stars would show
\vlos ~$\sim 0 $ km s$^{-1}$, almost completely losing one of the 4 coordinates. In this case, \beacon~assigned
approximately 70\% of the stars to streams, but with nearly 25\% of them badly allocated. This latter scenario is
however a very extreme case and unlikely the case for Fornax since we in fact detect rotation. The observation from
this point of view is probably the worst case we can encounter, and would require proper motions measurements in order
to improve the results. Further tests of \beacon~using $N$-body simulations will be discussed in a follow-up paper.

\section{Discussion}\label{Cap:Discussion}

Stars in Fornax show a large variety of orbits which appear to follow certain kinematical patterns. Disentangling these
patterns is difficult with only partial information usually available to observers. \beacon~has demonstrated to be very
powerful in this task, being able to detect reliable chemo-kinematic patterns. Our results show that Fornax rotates,
and it has quite complex rotation patterns.

Fornax shows a non-negligible rotation signal ($\sim2\sigma$ over \vlos = 0), mostly around its optical minor
axis (Figure~\ref{fig:Rotation_Curve}). As has been already discussed in Sections~\ref{Subcap:Angular_momentum},~\ref{Subcap:Mass} and \ref{Cap:Consistency}, this is consistent with Fornax being a triaxial system, at least partially supported by rotation.

There exist several mechanisms that could produce a spheroidal galaxy from a progenitor galaxy with a different
shape. These include tidal stirring, star formation and supernova feedback or mergers between dwarf galaxies. We believe that a more precise age determination than the one achieved in this work is required
in order to derive strong conclusions related to the specific moments in which Fornax could have suffered from any external event, i.e. interaction with other systems.
However, we think it is illustrative to make an outline of the properties of Fornax that could be related to these interactions.

Tidal interactions with the MW could be responsible of reshaping a disk galaxy into a dSph.
These have been studied in detail by means of high resolution $N$-body simulations \citep{Mayer2001, Kliment2009,
Kazan2011, Lokas2011, Lokas2012}. In such controlled experiments, a primordial late type galaxy is placed on an orbit
around a MW-like galaxy and evolved for a significant fraction of the Hubble time. Results show that with tight enough
orbits a dIrr galaxy can be transformed into a dSph, conserving some signatures of this interaction. The final galaxy
typically has a triaxial (mostly prolate) shape with some remnant rotation around the minor axis. A detailed discussion
regarding the possible remaining effects of these interactions on the photometry of Fornax can be found in
\citet{delPino2015}.

Kinematic measurements provide complementary information about the past of the galaxy. The fact that some rotation
remains in Fornax may indicate a rotating progenitor for this galaxy. Figure~\ref{fig:Age_KLA} shows the Fornax
perigalacticon passages expected from the orbital parameters estimated using the presently available proper motion
measurements \citep[for further information see][section 7.1]{delPino2013}. There is no clear correlation between
perigalacticon passages and the dynamics of the galaxy, although our data do not cover stars distant
enough to study possible tidal tails or other dynamical signatures from tidal stirring. 

On the other hand, a drop in $|\mathbf{L}|$ can be noticed around the first perigalacticon passage, $\sim 9.8$ Gyr ago.
This passage coincides with the change in the angular momentum orientation from the one pointing along $+b$ to the
one pointing along $+a$.
Interestingly, at the second passage
($\sim 6.2$ Gyr ago), when the rotation of the galaxy is around $-b$, we observe an enhancement of $|\mathbf{L}|$.
Several scenarios could explain this effect, including the tidal stirring model. The orientation of the
angular momentum of Fornax with respect to the MW potential during the perigalacticon passages could have influenced the
orbits of the stars in Fornax. In fact, tidal effects are strongly dependent on the relative orientation of the angular
momentum of the stars in the galaxy with respect to the orbital momentum of the galaxy around its host \citep[see for
example][]{Lokas2015}. Tidal forces are much more effective in weakening the rotation streaming motions for dwarf
galaxies on prograde orbits. However, if a bar is present in the dwarf the host galaxy can also accelerate the
orbits of the stars in the dwarf depending on the orientation of the bar at pericentre. For more details about how tidal forces can modify the pattern speed of the bar, we refer the reader to Section 6 of \citet{Lokas2014a}.

Our results are qualitatively
compatible with a scenario in which a disky progenitor of Fornax was on a prograde orbit around the MW at its first
close fly-by, so that its angular momentum was reduced and its orientation changed. This first perigalacticon passage
may have also resulted in the morphological transformation into a prolate (bar-like) spheroid. At the second
perigalacticon, if the orientation of the bar with respect to the direction of the tidal force from the MW was
favourable the angular momentum could have been increased again. Indeed this could be the case for stars born around 6 Gyr whose
angular momentum is approximately orthogonal and prograde to the proper motion vector of the galaxy, $\theta_{\rm pm} = 169^\circ\pm8.5^\circ$
in the Fornax main axis system \citep{Piatek2007}. However, this scenario does not explain the sudden
change of the orientation of the angular momentum $\sim 8.7$ Gyr ago, from the one along
its positive minor main semi-axis ($+b$) to the one around its negative minor semi-axis ($-b$).
Still, taking into account other possible tidal stirring signatures from photometry we can not rule out possible tidal
interactions of Fornax with the MW in the past.

Another probable scenario of the evolution of Fornax could involve a merger between two or more stellar systems. This
possibility have been considered in several works \citep{Coleman2004, Yozin2012, Amorisco2012}. The basic
idea relies on two disk galaxies colliding, each one with a particular angular momentum. As a result, the orbits of
the stars are disturbed, showing more chaotic and random rotation patterns. After the merger the system would
have a spheroidal shape and a relatively high velocity dispersion. This scenario is also compatible with more than one
system being accreted and does not rule out the possibility of different kinds of galaxies colliding (e.g. a disk and a
spheroid).

Some signatures could remain in the galaxy after these events for several Gyr: strange rotation patterns, shell-like
structures, clumps or streams of stars, etc. Fornax appears to show some of these characteristics, making it very
plausible that a merger had played an important role in its evolution or even formation. These
features have already been discussed from the point of view of photometry in \citet{delPino2013,delPino2015}. In the
present work, we revisit the problem using the information provided by spectroscopy, i.e. \vlos ~and metallicity.

The existence of different rotation patterns as a function of the stellar age (see Figure~\ref{fig:Z_A}) may point to
an accretion of stellar material with different angular momentum at some time in the past. We have traced the preferred
rotation direction as a function of the stellar age and found that approximately between 10 and 7.5 Gyr ago the
rotating motion undergoes some fluctuations (see Figure~\ref{fig:Age_KLA}). These changes may not be
significant, since stars between 10 and 7.5 Gyr old show little streaming motion, i.e. low $|\mathbf{L}|$. After that,
the galaxy appears to steadily recover the preferred rotating direction around its minor axis. Several
scenarios could explain this behaviour, but probably the simplest and most straightforward is, indeed, a gas-rich merger
between two dwarfs. We therefore find further support for the scenario proposed in \citet{delPino2013,delPino2015}: a
major merger suffered by Fornax at $z\sim 1$. The violent event would have expelled a significant amount of gas which
would then be accreted on to the deeper potential well of the resulting galaxy. This scenario would explain the
conspicuous dynamical features shown by Fornax.

First, the gas from both progenitors could have had different angular momentum, resulting in a weaker angular
momentum of the mixed gas. This effect has been also observed during major mergers in the Illustris simulation
\citep{Genel2015}. The $\mathbf{L}$ shows significant fluctuations starting at $\sim 10$ Gyr ago, with the largest
change in its direction at $\sim 8$ Gyr ago. This would indicate some prolonged gravitational interaction between the
progenitors until they finally merged. The latter is consistent with the scenario proposed by \citet{Amorisco2012} in
which Fornax is the result of a merger of a bound pair of galaxies. These galaxies would have interacted
gravitationally, showing as a result the fluctuations observed in $\mathbf{L}$. This event would have also affected the
SFH of the galaxy. In the top panel of Figure~\ref{fig:Age_KLA} we show the SFH for the three observed fields defined
in \citet{delPino2013}: IC1, IC2 and OC. Some interesting correlations between the SFHs and the rotation direction can
be observed. The first one shows itself at about 10 Gyr ago, when we see a strong burst of star formation in the
outskirts of the galaxy. This was followed by the most intense burst of star formation, located in the central regions
of the galaxy (IC1) at $\sim 8$ Gyr ago.

Second, the galaxy could have experienced an inflow of the previously expelled gas. Since this gas would not have
had the same angular momentum as the galaxy, its accretion may have caused the chaotic rotation patterns observed
in the younger populations. This gas would also be responsible of the shell-like structure found by
\citet{Coleman2004}.

This scenario would be consistent with comprehensive cosmological $\Lambda$CDM simulations of Local Group-like
environments in which only very few mergers between dwarf galaxies happen later than $z\sim 1$ \citep{Kliment2010}. It
is also consistent with the bound pair merger scenario proposed by \citet{Amorisco2012} and with the chemical
pre-enrichment of the inner shell structure \citep{Olszewski2006}.

\section{Conclusions}\label{Cap:Conclusions}

We have detected rotating streams of stars in Fornax. This indicates that Fornax is a rather complex
system with various rotating components and that its spheroidal shape is the superposition of stellar components with
distinct kinematics. \beacon~has proved to be a useful tool for detecting these kinematic components.

Fornax rotates mainly around its minor optical axis with an average velocity of $\sim 12$ km s$^{-1}$. This is
consistent with Fornax being a triaxial system partially supported by rotation. However, its detailed
chemo-kinematics is complex, showing multiple rotation patterns depending on the metallicity of the stars.
Stars with metallicities \feh$\sim - 1.1$ have more chaotic patterns, resulting in nearly complete lack of global
rotation. On the other hand, more metal rich stars show strong rotation signal, but in the opposite direction
with respect to the dominant rotation.

We have derived the age for the groups of stars using their metallicity distributions. Our results show that stars with
ages around 8 Gyr ($z\sim1$) change the direction of their rotation. We conclude that this change, together
with the strong spatial gradients observed in the SFH as a function of radius \citep{delPino2013}, and the more chaotic
angular momenta shown by the younger populations are closely related. We propose a merger occurring at $z \sim 1$ as
a plausible scenario for explaining all these features. This scenario is consistent with the results from photometry
presented in \citet{delPino2015}.

We have also investigated the possible effects of tidal interactions of Fornax with the MW. Our results suggest that
some interactions between Fornax and the MW could have occurred during the perigalacticons. The first one, that
probably occurred $\sim 9.8$ Gyr ago coincides with an almost complete loss of rotation in the galaxy until $\sim 7.5$
Gyr ago. Moreover, the enhancement observed in the rotation signal during the second passage could also be caused by
tides if the orientation of the stellar component was favourable. On the other hand, according to models by 
\citet{Mayer2010}, Fornax should have had its first perigalacticon passage
after $z=1$ in order to have such an extended star formation history. This, together with the lack of data at large enough distances from the galaxy, does not allow
us to detect direct signatures of such interactions, e.g. tidal tails. Therefore, the relevance of these interactions
for the evolution of Fornax remains unclear.

\section*{Acknowledgments}

The authors thank the referee, Dr. Mario Mateo for his thorough review
and  comments. We are grateful to Dr. G. Battaglia and Dr. E. Kirby for generously providing their data for this project, and to Dr. M. Monelli and Dr. S. Fouquet
for their comments and suggestions which have helped to improve the manuscript. AdP also thanks S. Bertran de Lis for her support
during this project. This research was partially supported by the IAC (grant 310394) and the Education and Science Ministry of Spain
(grants AYA2010-16717, and AYA2013-42781-P). Further support has been provided by the Polish National Science Centre
under grant 2013/10/A/ST9/00023.

\label{lastpage}
\end{document}